\documentclass[aip,twocolumn,reprint,showpacs, groupedaddress]{revtex4-2}
\usepackage{amsmath}
\usepackage[latin9]{inputenc}
\usepackage{graphicx}
\usepackage{hyperref}
\usepackage{graphicx}
\usepackage[color= blue!20]{todonotes}
\usepackage[normalem]{ulem}
\usepackage{color}
\usepackage{ulem}
\usepackage{xcolor}
\hypersetup{
	colorlinks,
	linkcolor={red!70!blue},
	citecolor={red!70!blue},
	urlcolor={red!70!blue}
}
\DeclareMathOperator{\atan2}{atan2}

\begin{document}

\title{Niobium Quantum Interference Microwave Circuits \\ with Monolithic Three-Dimensional (3D) Nanobridge Junctions}

\author{Kevin~Uhl}\email{kevin.uhl@pit.uni-tuebingen.de}
\affiliation{Physikalisches Institut, Center for Quantum Science (CQ) and LISA$^+$, Universit\"at T\"ubingen, 72076 T\"ubingen, Germany}
\author{Daniel~Hackenbeck}
\affiliation{Physikalisches Institut, Center for Quantum Science (CQ) and LISA$^+$, Universit\"at T\"ubingen, 72076 T\"ubingen, Germany}
\author{Janis~Peter}
\affiliation{Physikalisches Institut, Center for Quantum Science (CQ) and LISA$^+$, Universit\"at T\"ubingen, 72076 T\"ubingen, Germany}
\author{Reinhold~Kleiner}
\affiliation{Physikalisches Institut, Center for Quantum Science (CQ) and LISA$^+$, Universit\"at T\"ubingen, 72076 T\"ubingen, Germany}
\author{Dieter~Koelle}
\affiliation{Physikalisches Institut, Center for Quantum Science (CQ) and LISA$^+$, Universit\"at T\"ubingen, 72076 T\"ubingen, Germany}
\author{Daniel~Bothner}\email{daniel.bothner@uni-tuebingen.de}
\affiliation{Physikalisches Institut, Center for Quantum Science (CQ) and LISA$^+$, Universit\"at T\"ubingen, 72076 T\"ubingen, Germany}

\begin{abstract}
Nonlinear microwave circuits are key elements for many groundbreaking research directions and technologies, such as quantum computation and quantum sensing.
The majority of microwave circuits with Josephson nonlinearities to date is based on aluminum thin films, and therefore they are severely restricted in their operation range regarding temperatures and external magnetic fields.
Here, we present the realization of superconducting niobium microwave resonators with integrated, three-dimensional (3D) nanobridge-based superconducting quantum interference devices.
The 3D nanobridges (constriction weak links) are monolithically patterned into pre-fabricated microwave LC circuits using neon ion beam milling, and the resulting quantum interference circuits show frequency tunabilities, flux responsivities and Kerr nonlinearities on par with comparable aluminum nanobridge devices, but with the perspective of a much larger operation parameter regime.
Our results reveal great potential for application of these circuits in hybrid systems with e.g. magnons and spin ensembles or in flux-mediated optomechanics.

\end{abstract}

\maketitle
\let\oldaddcontentsline\addcontentsline
\renewcommand{\addcontentsline}[3]{}

\subsection*{Introduction}
\vspace{-2mm}

Superconducting microwave circuits with integrated Josephson junctions (JJs) and superconducting quantum interference devices (SQUIDs) have led to groundbreaking experimental and technological developments in recent decades. 
Both, single JJs and SQUIDs constitute a flexible and designable Josephson or Kerr nonlinearity, while a SQUID additionally provides in-situ tunability of the resonance frequency by external magnetic flux.
Circuits with large nonlinearities originating from the Josephson element form artificial atoms and qubits \cite{Clarke08, You11}, which have been used for spectacular experiments in circuit quantum electrodynamics \cite{Blais21} and quantum information processing \cite{Arute19}.
Frequency-tunable devices with a small nonlinearity are highly relevant for quantum-limited Josephson parametric amplifiers \cite{CastellanosBeltran08, Bergeal10, Macklin15}, tunable microwave cavities for hybrid systems with spin ensembles and magnons, for dispersive SQUID magnetometry \cite{Hatridge11, LevensonFalk16}, for photon-pressure systems \cite{Johansson14, Eichler18, Bothner21} and for microwave optomechanics \cite{Shevchuk17, Rodrigues19, Schmidt20, Zoepfl23}.
In many of these currently active research fields, such as flux-mediated optomechanics, hybrid quantum devices with magnonic oscillators and dispersive SQUID magnetometry, it is highly desirable to have frequency-tunable microwave circuits with small nonlinearity, high magnetic field tolerance and (in some cases) a critical temperature significantly above that of aluminum.
The vast majority of frequency-tunable and nonlinear circuits, however, uses Josephson junctions and SQUIDs made of aluminum thin films \cite{Nakamura99, PalaciosLaloy08}, a superconducting material with a critical magnetic field of only $B_\mathrm{c} \sim 10 - 100\,$mT (depending on material properties) and a critical temperature $T_\mathrm{c} = 1.2\,$K.
An approach that could fulfil the aforementioned wishlist is the implementation of microwave circuits made of niobium \cite{deGraaf12}, niobium alloys \cite{Samkharadze16, Zollitsch19} or even a high-$T_\mathrm{c}$ superconductor such as YBCO \cite{Ghirri15, Roitman23, Uhl23} with high critical current density and high-field compatible Josephson elements such as nano-constrictions \cite{Romans11, Kennedy19, Xu22}.  
In most efforts so far, however, it has been proven difficult to obtain large-tunability constriction-junction SQUIDs made of these materials, both in direct current (DC) operation mode and in the microwave domain \cite{Kennedy19, Bouchiat01, Hasselbach02, Mitchell12, Wang17}.
Here, we report the realization of niobium superconducting quantum interference microwave circuits based on focused neon-ion-beam patterned monolithic 3D nanobridge junctions.
Although the SQUIDs in our devices have a large effective area of $\sim 72\,$\textmu m$^2$, we achieve smaller screening parameters than previous 2D niobium nanobridge SQUID circuits with much smaller loops \cite{Kennedy19}.
A small screening parameter is an important prerequisite for stable flux-tunability of the circuit resonance frequency and large flux responsivities \cite{Pogorzalek17}.
In addition, we characterize our nanobridge quantum interference circuits at varying temperatures in the regime $2.4\,$K$\,<T_\mathrm{s}<3.4\,$K and demonstrate that they have only a small Kerr nonlinearity of $|\mathcal{K}|/2\pi \lesssim 10\,$kHz, ideal for large dynamic range applications.
Our devices and results show great potential for dispersive SQUID magnetometry, hybrid systems with spin ensembles, magnons or cold atoms and for flux-mediated optomechanics.

\begin{figure*}
	\centerline{\includegraphics[clip=True,width=0.85\textwidth]{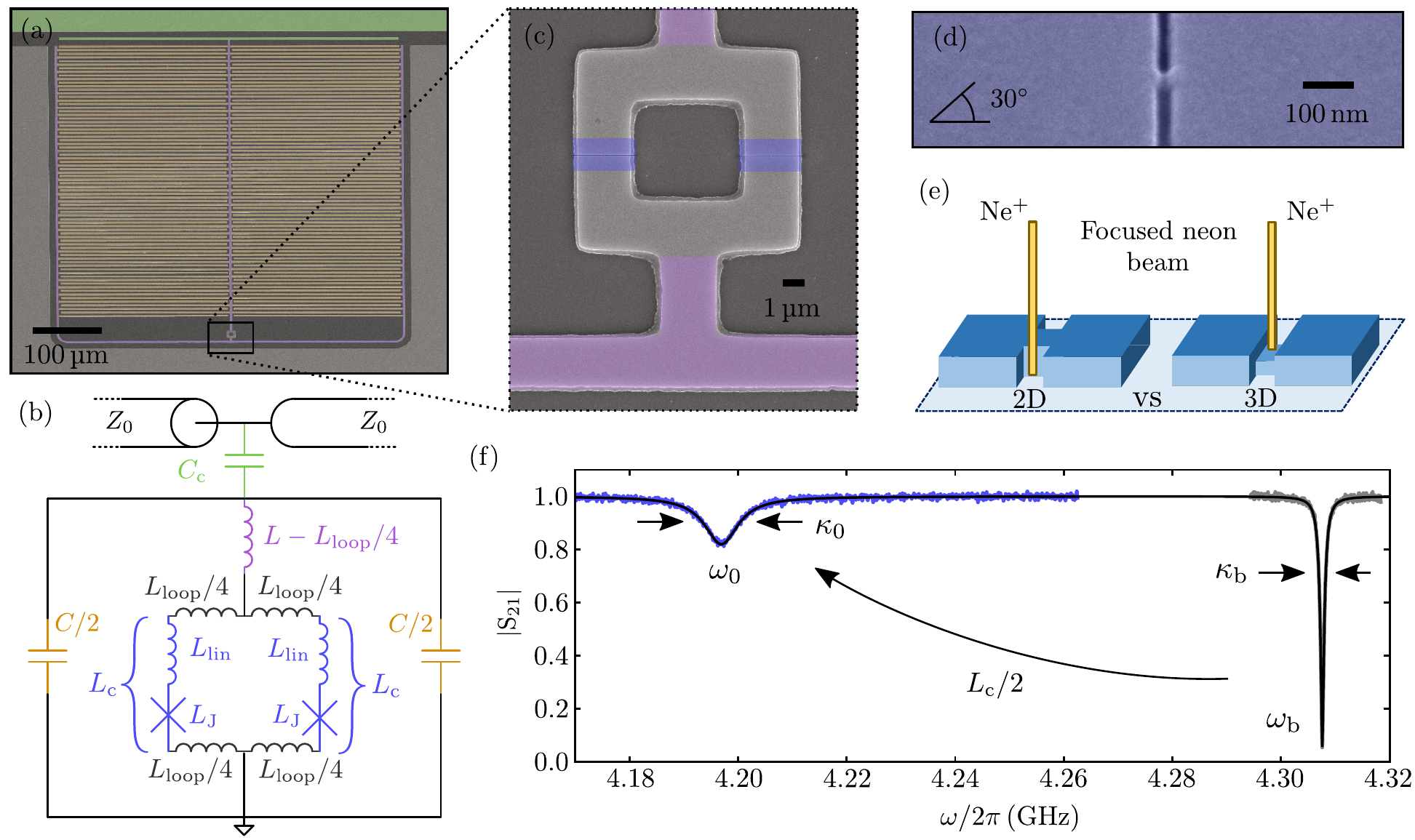}}
	\caption{\textsf{\textbf{A niobium-based quantum interference microwave circuit with monolithic three-dimensional nanobridge junctions.} (a) False-colour optical micrograph and (b) equivalent circuit of a typical device. The circuit main inductance (purple) is modeled by a linear inductor $L-L_\mathrm{loop}/4$, the two interdigitated capacitors (IDCs, orange) by a total capacitance $C$. Each of the two IDCs has $N_\mathrm{idc}$ (here $N_\mathrm{idc} = 46$) fingers with a length of $l = 250\,$\textmu m and a width of $W = 3\,$\textmu m. In the center of the circuit is a loop structure for the superconducting quantum interference device (SQUID). The square-shaped loop has a total loop inductance $L_\mathrm{loop}\approx 17\,$pH and the nanobridges (blue) have a constriction inductance $L_\mathrm{c}$ (only $\neq 0$ after junction patterning). The resonant circuit is capacitively coupled to a coplanar waveguide transmission line with a characteristic impedance of $Z_0 \approx 50\,\Omega$ by means of a coupling capacitance $C_\mathrm{c}$ (coupling elements green). (c) False-color scanning electron microscopy (SEM) image of the loop after constriction cutting, (d) zoom of a 3D constriction after cutting, taken with a SEM tilt angle of 30°. In (a), (c), (d) niobium is bright/colored, silicon substrate dark gray. Panel (e) schematically illustrates the nano-constriction fabrication. For the 2D constrictions, two narrow slits are patterned into each of the SQUID arms by a focused neon ion beam; for the 3D constrictions, the nanobridges are additionally thinned down from the top by the neon beam. (f) Transmission $|S_\mathrm{21}|$ of one of the circuits (device 3D$_1$) at  $T_\mathrm{s} = 2.5\,$K before (grey) and after (blue) the constriction cutting, black lines are fits to the data. Before the junction cutting the circuit has a resonance frequency $\omega_\mathrm{b}$ and a linewidth $\kappa_\mathrm{b}$, after the cutting $\omega_0$ and $\kappa_0$, respectively. Values can be found in the main text. From the shift in resonance frequency by the cutting, we determine the total additional inductance of the constrictions $L_\mathrm{c}/2$.}}
	\label{fig:Figure1}
\end{figure*}

\subsection*{Devices}
\vspace{-2mm}
Our devices are lumped element microwave circuits, patterned from a $d_\mathrm{Nb} = 90\,$nm thick layer of magnetron sputtered niobium on top of a high-resistivity silicon substrate with thickness $d_\mathrm{Si} = 500\,$\textmu m.
They consist of two interdigitated capacitors (IDC) combined in parallel and several linear inductors, and they are capacitively side-coupled to a $Z_0 \approx 50\,\Omega$ coplanar waveguide transmission line by means of a coupling capacitance $C_\mathrm{c}$ for driving and readout.
One of the devices and its circuit equivalent are shown in Fig.~\ref{fig:Figure1}(a) and (b), respectively.
The width of all lines (fingers and inductor wires) and the gaps in between two adjacent IDC fingers is $W = 3\,$\textmu m.
At the connection point between the capacitors and the inductor wires, a square-shaped loop with an effective area of $\sim 8.5\times 8.5\,$\textmu m$^2$ (hole size $6\times 6\,$\textmu m$^2$) is embedded into the circuit, which forms the SQUID once the nano-constrictions are introduced, cf. Fig.~\ref{fig:Figure1}(c, d).
After patterning the circuit itself by means of optical lithography and reactive ion etching using SF$_6$, the nano-constrictions are fabricated into the center of the two loop arms using a focused neon ion beam.
For the simplest constrictions, we cut a narrow $20\,$nm wide slot from both sides into each of the two loop arms, leaving only a $40\,$nm wide constriction in the center of each arm.
This type of constriction (thickness equal for the leads and the constriction) has been referred to as 2D constriction and has been implemented for both, DC and microwave SQUIDs \cite{Bouchiat01, Hasselbach02, Kennedy19}.
It has also been demonstrated, however, that 3D versions, i.e., constrictions that are thinner than the superconducting leads connected to them, can have superior properties such as less skewed current-phase-relations and smaller critical currents \cite{Bouchiat01, Vijay10, Chen16}.
Implementing 3D constrictions allows for keeping the circuit film thickness large, i.e., the circuit and loop kinetic inductance small, while at the same time getting a critical junction current $I_0 \sim 10\,$\textmu A, a highly desirable range for simultaneously achieving a large frequency tunability and a small circuit nonlinearity.
For the implementation of the 3D versions, we therefore modify our ion beam scan pattern in a way that the constriction is thinned down during the cutting procedure, cf. Fig.~\ref{fig:Figure1}(e), more details can be found in Supplementary Note~I.
Such a monolithic approach for the generation of 3D nanobridges circumvents some of the challenges and possible problems of previously implemented multi-layer deposition processes, such as guaranteeing good galvanic contact between the layers or dealing with thin additional edges at the bottom of the microwave structures \cite{Vijay10, Chen16}.
Furthermore, it allows for the unique opportunity to characterize one and the same microwave circuit both without and with the junctions, i.e., one can determine experimentally the impact of the junctions on the circuit properties.
We combine several LC circuits on a single coplanar waveguide feedline, more specifically four circuits with a SQUID and three circuits without a SQUID for reference.
The base circuits only differ in the number of fingers in the IDCs and in the corresponding resonance frequencies between $3$ and $7\,$GHz.
We present data for three of the SQUID resonators with three different constriction types; one resonator has 2D constrictions (junction thickness $d_\mathrm{JJ} = 90\,$nm) and two resonators have 3D constrictions with $d_\mathrm{JJ} \approx 30\,$nm (3D$_1$) and with $d_\mathrm{JJ} \approx 20\,$nm (3D$_2$), respectively (thicknesses are estimated from the neon dose).
The total chip is $10\times 10\,$mm$^2$ large and is mounted into a printed circuit board (PCB), to which both the ground planes and the coplanar waveguide feedlines are connected through wirebonds.
Chip and PCB are placed in a radiation-tight copper housing and the package -- including a magnet coil fixed to the box -- is mounted into the vacuum chamber of a dipstick that can be inserted into a liquid Helium cryostat.
The cryostat allows for a high-stability temperature control in the range $2.4\,$K$<T<7.5\,$K by a combination of pumping on the liquid helium container and a feedback loop using a temperature diode and a heating resistor in the vacuum compartment where the sample is mounted.
The sample box including the magnet is placed additionally into a cryoperm magnetic shield and the whole cryostat is packed into a double-layer room temperature mu-metal shield.
The microwave input line is strongly attenuated with $\sim 30\,$dB to equilibrate the incoming noise to the sample temperature and the output line is connected to a cryogenic high-electron-mobility-transistor amplifier.
More details on the experimental setup are given in Supplementary Notes~II and III.

\begin{figure*}
	\centerline{\includegraphics[ clip=True,width=0.97\textwidth]{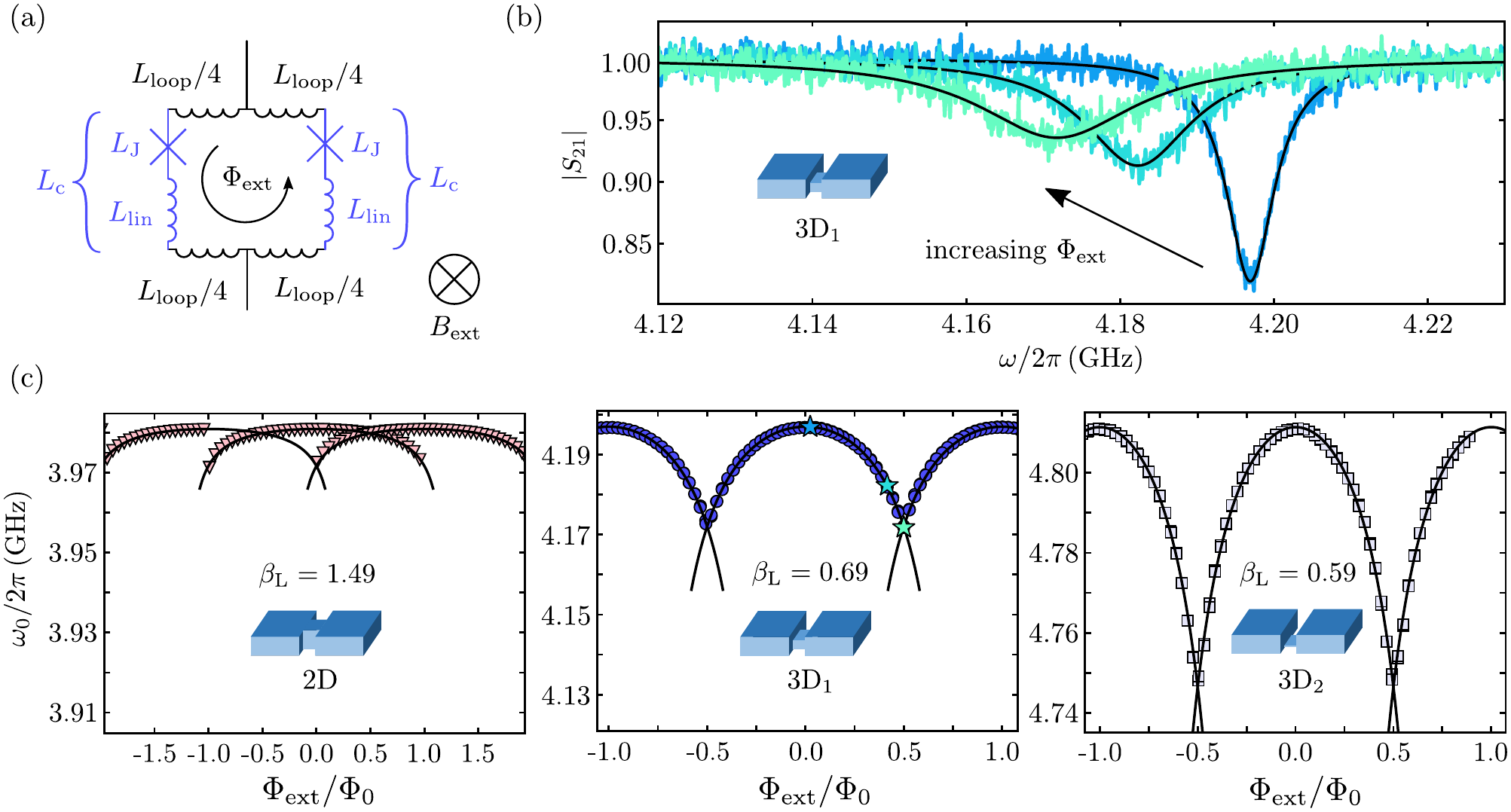}}
	\caption{\textsf{\textbf{Flux-tuning the resonance frequency of niobium quantum interference circuits with 2D and 3D constriction junctions.} (a) Circuit equivalent of the SQUID with a linear loop inductance $L_\mathrm{loop}/2$ in each arm and a constriction inductance $L_\mathrm{c}$, modeled by a linear contribution $L_\mathrm{lin}$ and a sinusoidal Josephson contribution $L_\mathrm{J}$. We apply an external magnetic field $B_\mathrm{ext}$ perpendicular to the SQUID loop to change the inductances $L_\mathrm{J}(\Phi_\mathrm{ext})$ with the externally applied magnetic flux $\Phi_\mathrm{ext}$ and with it the resonance frequency of the circuit. (b) Transmission response $|S_{21}|$ of the 3D$_1$ constriction SQUID circuit for three different external fluxes $\Phi_\mathrm{ext}$, $T_\mathrm{s} = 2.5\,$K. With increasing $\Phi_\mathrm{ext}$, the resonance is shifting towards lower frequencies, indicating an increase of the constriction inductance by flux. Colored noisy lines are data and black smooth lines are fits. The flux values are $\Phi_\mathrm{ext}/\Phi_0 = 0, 0.4, 0.5$. From the fits, we extract the resonance frequency $\omega_0(\Phi_\mathrm{ext})$, which is shown as a function of flux in panel (c) for the three different circuits with three different constrictions. Insets show sketches of the junction type. Symbols are data, lines are fits from which we extract the screening parameter $\beta_\mathrm{L}$. Left: 2D constriction, thickness $\sim 90\,$nm. Middle: 3D constriction 3D$_1$, thickness $\sim 40\,$nm, star-shaped data points correspond to data sets in (b). Right: 3D constriction 3D$_2$, thickness $\sim 20\,$nm. With decreasing thickness of the constriction, the tuning range gets larger and screening parameter $\beta_\mathrm{L}$ and flux hysteresis (overlap of adjacent flux archs) decrease.}}
	\label{fig:Figure2}
\end{figure*}

\subsection*{Results - Impact of junction cutting}
\vspace{-2mm}

As first step in our device characterization, we measure the transmission coefficient $S_{21}$ with a vector network analyzer (VNA) once before the constriction patterning and once after.
%
%
In Fig.~\ref{fig:Figure1}(f), the transmission $|S_{21}|$ at $2.5\,$K for the device 3D$_1$ is shown in direct comparison for both cases.
From the fits, we obtain the resonance frequencies $\omega_\mathrm{b} = 2\pi\cdot 4.308\,$GHz (no constrictions) and $\omega_\mathrm{0} = 2\pi\cdot 4.197\,$GHz (with constrictions) and therefore can calculate the single-constriction inductance from the constriction-induced frequency shift via
\begin{equation}
	\omega_\mathrm{0} = \frac{\omega_\mathrm{b}}{\sqrt{1 + \frac{L_\mathrm{c}}{2L}}}.
\end{equation}
Using $L = 568\,$pH as obtained from a combination of measuring the temperature-dependence of the resonance frequency and numerical simulations with the software package 3D-MLSI \cite{Khapaev01}, cf. Supplementary Note~IV for all device parameters, we get $L_\mathrm{c}^\mathrm{3D1} = 61\,$pH for the device 3D$_1$.
Additionally, we extract the internal (subscript 'i') and external (subscript 'e') linewidths from the fit before and after nanobridge patterning and obtain $\kappa_\mathrm{i, b} = 2\pi \cdot 73\,$kHz, $\kappa_\mathrm{e, b} = 2\pi\cdot 1.2\,$MHz without and $\kappa_\mathrm{i} = 2\pi \cdot 6.5\,$MHz, $\kappa_\mathrm{e} = 2\pi\cdot 1.4\,$MHz with the constriction junctions.
Here $\kappa_\mathrm{b} = \kappa_\mathrm{i, b} + \kappa_\mathrm{e, b}$ and $\kappa_0 = \kappa_\mathrm{i} + \kappa_\mathrm{e}$.
A change in the external linewidth is most likely due to a slightly different input impedance of the feedline from the resonator perspective before and after the junction cutting and could either be related to a frequency-dependent feedline impedance ($\omega_0 < \omega_\mathrm{b}$) or to a change of impedance induced by taking the sample out of the PCB for the cutting and mounting/wirebonding it back into the box afterwards.
The considerable increase of the internal linewidth indicates that cutting the junction has introduced an additional loss channel and we believe it is related to an increased quasiparticle density inside the constriction.
First, it has been observed that ion-milled constrictions have a reduced critical temperature compared to the rest of the niobium film \cite{Hazra15, DeSimoni20, Qing21, Wyss22}, which locally decreases the superconducting gap and increases the intrinsic thermal quasiparticle density, cf. also the later discussion of the temperature dependence of the devices.
Based on this reduced gap, the local quasiparticle density could even be further increased, since the constriction with the reduced gap might act as a potential well or trap for thermal quasiparticles from the leads, similar to what has been observed in aluminum constrictions or vortex cores \cite{LevensonFalk14, Nsanzineza14}.
We note that the increase of losses could also partly be related to generating some normal-conducting niobium at the surface or at the edges of the constriction by the neon ions.
To illuminate the exact loss mechanisms in detail, however, further and dedicated experiments will be necessary in the future.
By performing completely analogous experiments and data analyses for the circuits 2D and 3D$_2$, we extract the corresponding constriction inductances to be $L_\mathrm{c}^\mathrm{2D} = 8\,$pH and $L_\mathrm{c}^\mathrm{3D2} = 103\,$pH.
More details regarding the three circuits and their basic parameters $L$, $C$, $L_\mathrm{loop}$, $C_\mathrm{c}$ and $\kappa_\mathrm{i/e}$ can be found in Supplementary Note~IV.

\begin{figure}
	\centerline{\includegraphics[clip=True, scale=0.75]{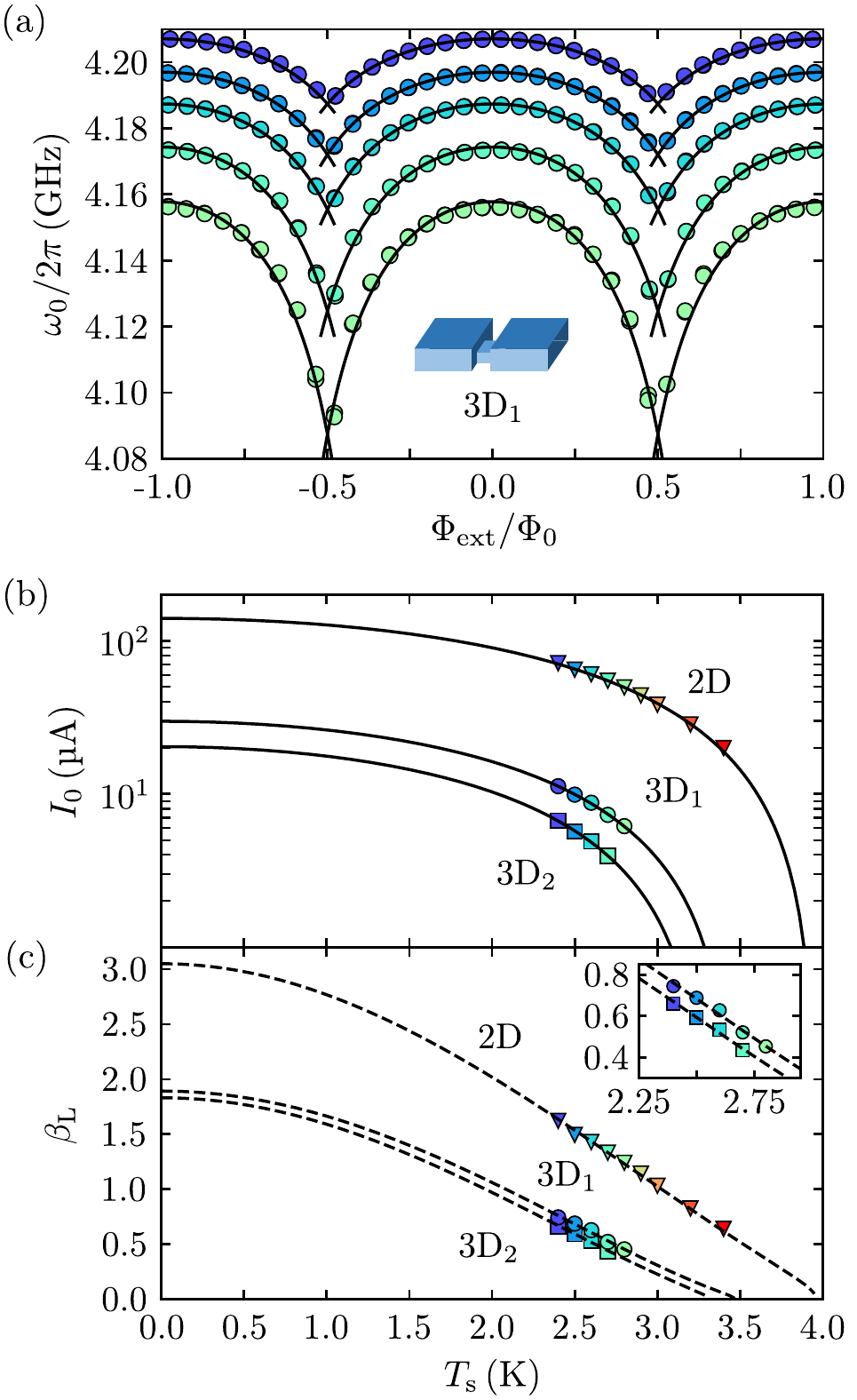}}
	\caption{\textsf{\textbf{Temperature dependence of circuit and SQUID.} (a) Flux-tuning curve $\omega_0(\Phi_\mathrm{ext})$ of device 3D$_1$ for several different sample temperatures $T_\mathrm{s} = 2.4-2.8\,$K in steps of $0.1\,$K. Top curve: lowest temperature. Bottom curve: highest temperature. Circles are data, lines are fits with Eqs.~(\ref{eqn:w0}) and (\ref{eqn:flux}). From the fits we extract the critical current of the constriction $I_0$ and the screening parameter $\beta_\mathrm{L}$, the obtained values for both are shown as circles in (b) and (c), respectively. (b) Critical current $I_0$ vs sample temperature $T_\mathrm{s}$ for all three constriction SQUID cavities as extracted from the corresponding flux-tuning curves. Symbols are data, lines are fits. From the theoretical fit curves (see main text), we can extrapolate to the critical current at mK temperatures and obtain the critical temperatures of the constrictions $T_\mathrm{cc}^\mathrm{2D} = 3.96\,$K, $T_\mathrm{cc}^\mathrm{3D1} = 3.47\,$K and $T_\mathrm{cc}^\mathrm{3D2} = 3.31\,$K. In combination with the temperature dependence of $L_\mathrm{loop}$ and $L_\mathrm{lin}$, we can calculate theoretical lines for $\beta_\mathrm{L}(T_\mathrm{s})$ as shown in panel (c) for all three circuits in comparison to the experimental data. Inset shows zoom-in to the data of the 3D samples.}} 
	\label{fig:Figure3}
\end{figure}

\subsection*{Results - Flux dependence}
\vspace{-2mm}

In order to learn more about the nature of the constrictions and how our SQUID circuits perform in terms of frequency-tunability, flux-responsivity and screening parameter, we apply an external magnetic field to the circuits that introduces magnetic flux $\Phi_\mathrm{ext}$ into the SQUIDs.
The constriction inductance $L_\mathrm{c}$ we obtained above is not necessarily a purely nonlinear Josephson inductance, but might have a linear contribution as well.
In many cases, nano-constrictions have been found to have forward-skewed sinusoidal current-phase relationships (CPRs) \cite{Hasselbach02, Vijay10, Likharev79, Gumann07, Troeman08} and such a skewed sine function can also be modeled approximately as a series combination of an ideal Josephson inductance $L_\mathrm{J}$ with sinusoidal CPR and a linear inductance $L_\mathrm{lin}$ \cite{Likharev79}, i.e., $L_\mathrm{c} = L_\mathrm{J} + L_\mathrm{lin}$, cf. Fig.~\ref{fig:Figure2}(a).
Here, the ideal Josephson inductance would be given by $L_\mathrm{J} = L_\mathrm{J0}/\cos{\varphi}$, where $\varphi$ is the phase difference across the junction, $L_\mathrm{J0} = \Phi_0\cdot(2\pi I_0)^{-1}$ and $I_0$ the critical current of the junction.
%
%
%
The Josephson phase $\varphi$ of each junction in a symmetric SQUID without bias current is related to the total flux $\Phi$ in the loop via $\varphi = \pi\Phi/\Phi_0$.
To change the magnetic flux through the loop, we sweep a DC current through the magnet coil attached to the backside of the chip housing, which generates a nearly homogeneous out-of-plane magnetic field at the position of the SQUIDs.
Figure~\ref{fig:Figure2}(b) shows the circuit response $|S_\mathrm{21}|$ of the 3D$_1$ SQUID circuit for several bias fluxes $\Phi_\mathrm{ext}$.
We observe that the resonance dip is moving to lower frequencies, i.e., that the resonance frequency is shifted downwards with flux, and that the depth of the dip decreases while the linewidth increases, at least as long as $\Phi_\mathrm{ext} < \Phi_0/2$ with the flux quantum $\Phi_0 \approx 2.068\cdot 10^{-15}\,$Tm$^2$. 
Over larger flux ranges, we in fact observe an oscillating behaviour of the resonance frequency $\omega_\mathrm{0}(\Phi_\mathrm{ext})$ with a periodicity of $\Phi_0$, reflecting fluxoid quantization in the SQUID loop, cf. Fig.~\ref{fig:Figure2}(c).
Very much as suggested by previous reports \cite{Vijay10, Pogorzalek17} and as intuitively expected, we observe that the resonance frequency tuning range (difference between maximum and minimum resonance frequency) gets larger with decreasing constriction thickness.
For the 2D junctions (left in Fig.~\ref{fig:Figure2}(c)), the total resonance frequency tuning range that we can achieve is only on the order of $\sim 10\,$MHz and the individual flux archs strongly overlap with a total observable width of $\sim 2\Phi_0$.
For the 3D$_2$ device (rightmost in Fig.~\ref{fig:Figure2}(c)), the tuning range is on the order of $\sim65\,$MHz and a flux hysteresis (two possible resonance frequencies for a single flux value as in the 2D circuit) is not observable in the data.
The 3D$_1$ circuit is somewhere in between, just as it is positioned in Fig.~\ref{fig:Figure2}(c). 
To quantitatively model the flux dependence of the resonance frequency and gather information about $L_\mathrm{J}$ and $L_\mathrm{lin}$, we consider a flux-dependent resonance frequency
\begin{equation}
	\omega_\mathrm{0}\left(\Phi_\mathrm{ext}\right) = \frac{\omega_\mathrm{b}}{\sqrt{1 + \frac{1}{2L} \left(L_\mathrm{lin} + \frac{L_\mathrm{J0}}{\cos{\left(\pi\frac{\Phi}{\Phi_0}\right)}} \right)}}.
	\label{eqn:w0}
\end{equation}
The relation between the total flux in the SQUID $\Phi$ and the external flux  $\Phi_\mathrm{ext}$ is given by
\begin{equation}
	\frac{\Phi}{\Phi_0} = \frac{\Phi_\mathrm{ext}}{\Phi_0} - \frac{\beta_\mathrm{L}}{2}\sin{\left(\pi\frac{\Phi}{\Phi_0}\right)}
	\label{eqn:flux}
\end{equation}
where
\begin{equation}
\beta_\mathrm{L} = \frac{2I_0\left(L_\mathrm{loop} + 2L_\mathrm{lin} \right)}{\Phi_0} = \frac{L_\mathrm{loop} + 2L_\mathrm{lin}}{\pi L_\mathrm{J0}}
\end{equation}
is the SQUID screening parameter.
The result of fitting the flux dependence of the resonance frequency with Eqs.~(\ref{eqn:w0}) and (\ref{eqn:flux}) is shown as lines in Fig.~\ref{fig:Figure2}(c) and shows good agreement with the experimental data for all three circuits.
%

%
%
%
The fit parameters we obtain for the single-junction sweetspot inductance $L_\mathrm{J0}$, the single-junction critical current $I_0$, the linear inductance contribution $L_\mathrm{lin}$ and the screening parameter $\beta_\mathrm{L}$ are summarized in Table.~\ref{tab:Table1}.
Interestingly, the extracted inductances do not show the somewhat expected tendency that $L_\mathrm{lin}/L_\mathrm{J0}$, representing the skewedness of the CPR, decreases with $d_\mathrm{JJ}$ and $I_0$.
As a consequence of the low critical current in the 3D$_2$ device, however, we obtain a small screening parameter $\beta_\mathrm{L} = 0.59$ despite our large SQUID loop and a maximum flux responsivity $\partial\omega_0/\partial\Phi_\mathrm{ext}\approx 2\pi\cdot 400\,$MHz$/\Phi_0$, on par with similar aluminum constriction devices \cite{Rodrigues22, Bothner22} and highly promising for applications in photon-pressure systems and flux-mediated optomechanics.

\begin{table}[h]
	\caption{\label{tab:params}\textsf{\textbf{Nanobridge and SQUID parameters for the three circuits.} Parameters according to the flux-tuning curve fits of Fig.~\ref{fig:Figure2}}(c) at $T_\mathrm{s} = 2.5\,$K. \\}
	\vspace{2mm}
	\begin{tabular}{ l l l l l}
		\hline
		Circuit ~~~ & $I_0\,$(\textmu A) ~~~ & $L_\mathrm{J0}\,$(pH) ~~~ & $L_\mathrm{lin}\,$(pH) ~~~ &  $\beta_\mathrm{\mathrm{L}}$ \\ \hline\hline
		2D & 65 & 5 & 3 & 1.49 \\ \hline
		3D$_1$ & 10 & 33 & 28 & 0.69 \\ \hline
		3D$_2$ & 6 & 58 & 45 & 0.59 \\ \hline
	\end{tabular}
\label{tab:Table1}
\end{table}

Regarding the increase of the linewidth with flux visible in the data of Fig.~\ref{fig:Figure2}(b), we believe it is related to a reduction of the superconducting gap with increasing current in the constriction \cite{Gumann07}, and the corresponding increase in local quasiparticle density, both by the reduction of the gap itself but also by trapping more quasiparticles from the leads.
Most likely, there are additional contributions due to internal and external low-frequency flux noise and thermal photon occupation of a nonlinear resonator \cite{PalaciosLaloy08}, which at several kelvin is on the order of $\overline{n}_\mathrm{th} \sim 10 - 20$ photons for a $4\,$GHz mode.

\subsection*{Results - Temperature dependence}
\begin{figure*}
	\centerline{\includegraphics[clip=True, width=0.95\textwidth]{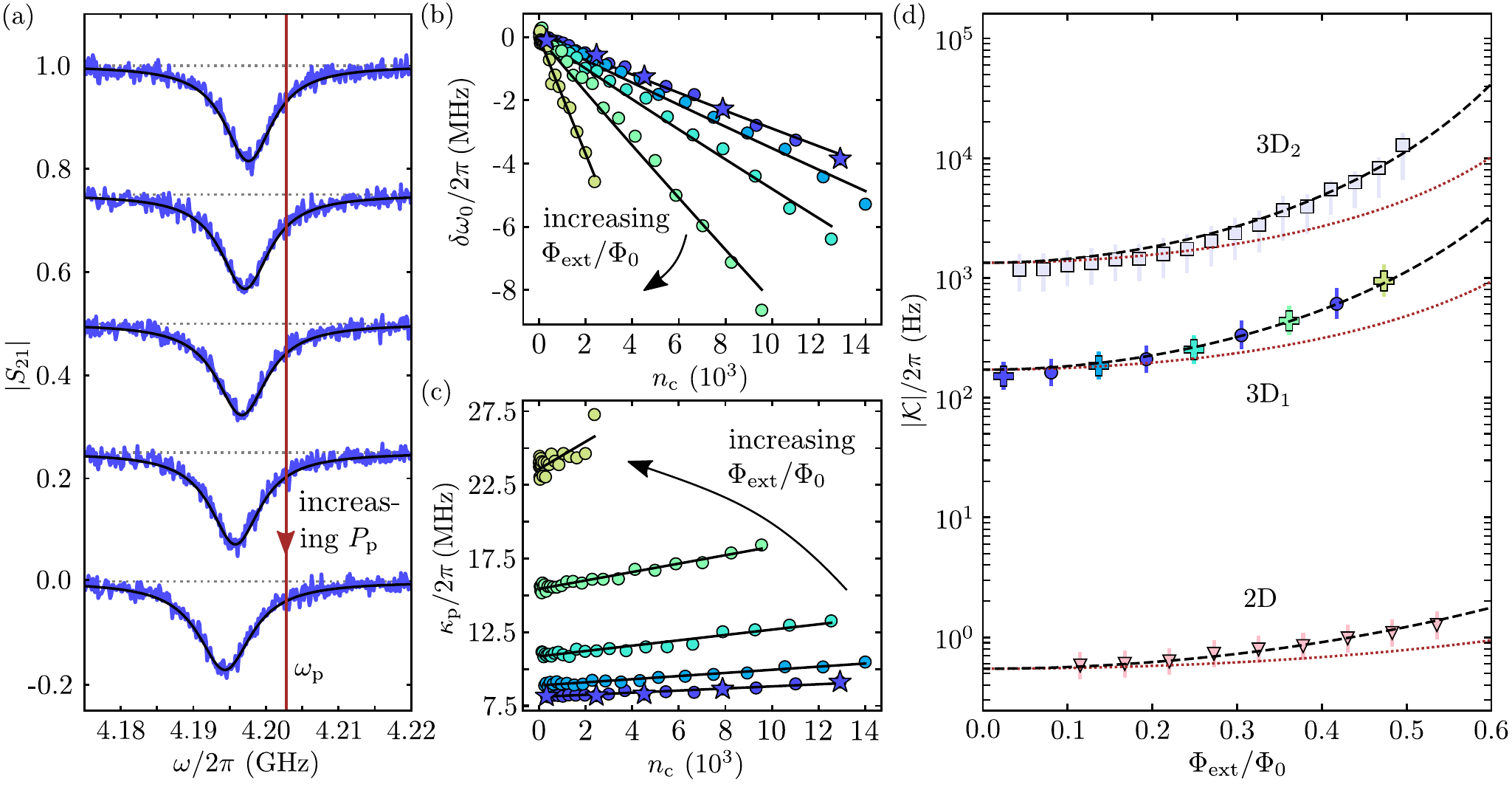}}
	\caption{\textsf{\textbf{Two-tone characterization of the SQUID-circuit Kerr nonlinearity.} (a) The SQUID circuit transmission response $S_{21}$ is probed with a weak microwave signal, while a strong microwave pump tone with fixed frequency $\omega_\mathrm{p}$ and variable power $P_\mathrm{p}$ is applied slightly blue-detuned from the cavity resonance $\omega_\mathrm{p} - \omega_0 = \varDelta_\mathrm{p} \sim \kappa$. With increasing $P_\mathrm{p}$, the dressed circuit resonance frequency is shifting towards lower frequencies and the linewidth increases. Five datsets for five different $P_\mathrm{p}$ are shown, subsequent datasets are offset by $-0.25$ for clarity. From fits (black lines) to the data (blue noisy lines) we calculate the resonance frequency shift $\delta\omega_0 = \omega_0(P_\mathrm{p})-\omega_0(0)$ and the linewidth $\kappa_\mathrm{p}$. The results are shown in panels (b) and (c) for different $\Phi_\mathrm{ext}$, temperature $T_\mathrm{s} = 2.5\,$K, sample 3D$_1$. Circles are data, lines are fits. Star symbols are data points that correspond to the five datasets shown in panel (a). From the fits, we determine the Kerr nonlinearity $\mathcal{K}(\Phi_\mathrm{ext})$. We perform this characterization for all three circuits, the result is shown in panel (c) vs external flux bias $\Phi_\mathrm{ext}/\Phi_0$. Symbols are data, dashed lines are theoretical curves based on the flux tuning curve and Eq.~(\ref{eqn:Kerr}) but without free parameters. Dotted red lines are theoretical curves without the correction factor arising from $\beta_\mathrm{L}\neq 0$, cf. main text and Supplementary Note X. Cross symbols in the data of 3D$_1$ correspond to the extracted values from the datasets shown in panels (b), (c). Error bars take into account uncertainties in the intracavity photon number $n_\mathrm{c}$, cf. Supplementary Note~IX.}}
	\label{fig:Figure4}
\end{figure*}

An interesting question when characterizing and operating superconducting microwave devices and SQUIDs at a temperature of several ten percent of the critical temperature $T_\mathrm{c}$ is how the properties depend on sample temperature $T_\mathrm{s}$ in that regime and if we can extrapolate to the properties at lower temperatures from that.
The most relevant parameters we are going to consider here are the cavity resonance frequency $\omega_0(T_\mathrm{s})$, the constriction critical current $I_0(T_\mathrm{s})$ and the SQUID screening parameter $\beta_\mathrm{L}(T_\mathrm{s})$ for all three circuits.
The main results are summarized in Fig.~\ref{fig:Figure3}.
We repeat the experiment of flux-tuning presented in the previous section for different sample temperatures $T_\mathrm{s}$.
From the transmission curves $S_{21}$ for varying external flux we extract $\omega_0(T_\mathrm{s}, \Phi_\mathrm{ext})$, cf. Fig.~\ref{fig:Figure3}(a) for a corresponding dataset of sample 3D$_1$.
Corresponding datasets for 2D and 3D$_2$ can be found in Supplementary Note~VIII.
For each temperature $T_\mathrm{s}$, we also measured $\omega_\mathrm{b}(T_\mathrm{s})$, so we have a reference resonance frequency from before neon irradiation, see also Supplementary Note~IV for the temperature dependence of the constriction-less 3D$_1$.
We observe in the resonance frequency tuning-curves that with decreasing temperature the zero-flux resonance frequency gets shifted to larger values, a clear indication for a reduction of the kinetic inductance both in the constrictions and in the rest of the circuit.
Furthermore, we find that the tuning range of the resonance frequency is growing with increasing temperature, indicating that the constriction inductance increases faster than the remaining circuit inductance and that the screening parameter $\beta_\mathrm{L}$ decreases, since $L_\mathrm{J0}$ is increasing faster with $T_\mathrm{s}$ than the effective loop inductance $L_\mathrm{loop} + 2L_\mathrm{lin}$.
For a quantification of these effects, we fit the flux-tuning data again with the same equations and procedure as described in the previous section.
As a result, we obtain for each sample $L_\mathrm{J0}(T_\mathrm{s})$ and $L_\mathrm{lin}(T_\mathrm{s})$, cf. Supplementary Note VIII, and from the former we calculate $I_0(T_\mathrm{s}) = \Phi_0\cdot\left[2\pi L_\mathrm{J0}(T_\mathrm{s})\right]^{-1}$.
The critical currents obtained from this are shown for all three circuits in panel (b).
We model the data with the theoretical Bardeen expression for the critical current of a constriction \cite{Bardeen62, DeSimoni20}
\begin{equation}
	I_0(T_\mathrm{s}) = I_\mathrm{c}\left[1 - \left( \frac{T_\mathrm{s}}{T_\mathrm{cc}}\right)^{2}  \right]^{3/2}
\end{equation}
and find as fitting parameters the critical current at zero temperature $I_\mathrm{c}$ as well as the constriction critical temperature $T_\mathrm{cc}$. 
As we already have speculated above, the critical temperature $T_\mathrm{cc}$ of the constrictions is considerably reduced compared to the niobium film to values between $T_\mathrm{cc}^\mathrm{2D} = 3.96\,$K and $T_\mathrm{cc}^\mathrm{3D2} = 3.31\,$K according to these fits.
Interestingly, similar $T_\mathrm{cc}$s have also been observed for elecron-beam-patterned niobium nanobridges with comparable critical currents \cite{DeSimoni20}, therefore the reduced transition temperature might actually not be related to an impact of the neon ions.
Since according to this fit our data are taken at $T_\mathrm{s} > 0.5 T_\mathrm{cc}$, we find that the critical currents still increase by a factor of two in the 2D constrictions and by about a factor of three in the 3D constrictions when approaching $T_\mathrm{s} \rightarrow 0$ and with respect to the smallest experimental temperature $T_\mathrm{s}^\mathrm{min} = 2.4\,$K.
It is also interesting to discuss the temperature dependence of $\beta_\mathrm{L}$ and its projected values in the mK regime, although this is a bit speculative due to the limited range of data available.
The experimental data shown in Fig.~\ref{fig:Figure4}(c) for all three circuits have been obtained from the flux-tuning fits.
For the 2D sample, the values for $\beta_\mathrm{L}$ are found to be between $0.6$ and $1.6$ in the measured regime and for the 3D samples between $0.4$ and $0.8$.
To model the temperature dependence we take into account the fit curves for $I_0(T_\mathrm{s})$ as shown in panel (b), the temperature dependence of the loop inductance $L_\mathrm{loop}(T_\mathrm{s})$ as discussed in Supplementary Note~IV and the temperature dependence of $L_\mathrm{lin}(T_\mathrm{s})$, that we obtain by a fit of the experimentally obtained values, cf. Supplementary Note~VIII.
We find curves which coincide very well with the experimental data in the measured range of $T_\mathrm{s}$ and which predict screening parameters for $T_\mathrm{s} \rightarrow 0$ saturating around $3.1$ for the 2D constrictions and around $1.8$ and $1.9$ for the 3D SQUIDs, respectively.
%

%
%
It seems that the non-sinusoidal CPR of the constrictions is currently the main limiting factor for $\beta_\mathrm{L}$ in the 3D samples, while in the 2D sample $2L_\mathrm{lin}^\mathrm{2D} \approx 6\,$pH and the screening parameter is limited by the actual $L_\mathrm{loop}$.
It will be interesting to see in future experiments in the mK temperature regime if these predictions are valid or if so far neglected effects will emerge and lead to a different behaviour than expected.
With lower-temperature experimental possibilities, it will also be interesting to further reduce the thickness and critical current of the 3D junctions, which in the current setup with $T_\mathrm{s}^\mathrm{min} \approx 2.4\,$K was not possible as can be seen from the very limited temperature range already accessible in the existing 3D devices with $T_\mathrm{cc} < 3.5\,$K.

\subsection*{Results - Kerr nonlinearity of the circuits}
\vspace{-2mm}
As final experiment we determine a very important parameter of Josephson-based microwave circuits -- their Kerr constant $\mathcal{K}$, also called anharmonicity or Kerr nonlinearity, which is equivalent to the circuit resonance frequency shift per intracavity photon.
For many applications a small Kerr nonlinearity is highly desired, as it increases the dynamic range or maximum intracavity photon number of the device, respectively.
This is important for instance in parametric amplifiers \cite{CastellanosBeltran08, Bergeal10, Macklin15, Frattini18} and in cavity-based detection techniques such as dispersive SQUID readout \cite{Hatridge11, LevensonFalk16}, SQUID optomechanics \cite{Rodrigues19, Zoepfl20} and photon-pressure sensing \cite{Eichler18, Bothner21}, where the signal of interest is proportional to the intracavity photon number $n_\mathrm{c}$ and therefore profits from high-power detection tones.
The origin of the nonlinearity in our SQUID circuits is the nonlinear inductance of the nano-constrictions.
In order to access $\mathcal{K}$ experimentally, we implement a two-tone protocol, cf. e.g. Refs.~\cite{Frattini18, FaniSani21}, and measure the equivalent of the AC Stark shift in the driven circuits.
The first microwave tone of the two-tone scheme is a fixed-frequency pump tone with variable power $P_\mathrm{p}$ and a frequency $\omega_\mathrm{p}$ slightly blue-detuned from the undriven cavity resonance $\omega_\mathrm{p} = \omega_0 + \varDelta_\mathrm{p}$ with $\varDelta_\mathrm{p} \sim \kappa$.
For each pump power, we then measure the pump-dressed device transmission $S_\mathrm{21}$ with a small probe tone, cf. Fig.~\ref{fig:Figure4}(a).
What we find qualitatively in this experiment is that with increasing pump power, the dressed circuit resonance frequency is shifting towards lower frequencies and that the internal linewidth of the mode is increasing, cf. Fig.~\ref{fig:Figure4}(a).
For a quantitative analysis, we fit each pump-dressed resonance with a linear cavity response (cf. Supplementary Note~V) for $S_{21}$, from which we extract the pump-shifted resonance frequency $\omega_0'$ and the pump-broadened total linewidth $\kappa_\mathrm{p}$.
To model the circuit and the results and to extract $\mathcal{K}$, we use the equation of motion for the complex intracavity field $\alpha$
\begin{equation}
	\dot{\alpha} = \left[ i\left(\omega_0 + \mathcal{K}|\alpha|^2\right) - \frac{\kappa_0 + \kappa_\mathrm{nl}|\alpha|^2}{2} \right]\alpha + i\sqrt{\frac{\kappa_\mathrm{e}}{2}}S_\mathrm{in}
\end{equation}
with a nonlinear damping parameter $\kappa_\mathrm{nl}$, the total input field $S_\mathrm{in}$ and with a normalization such that $|\alpha|^2 = n_\mathrm{c}$ is the total intracircuit photon number.
In the linearized two-tone regime (pump power $\gg$ probe power, details see Supplementary Note~VI), we find for the pump-broadened linewidth and the pump-induced frequency shift $\delta\omega_0 = \omega_0 - \omega_0'$ the relations
\begin{eqnarray}
	\kappa_\mathrm{p} & = & \kappa_0 + 2\kappa_\mathrm{nl}n_\mathrm{c} 	\label{eqn:Kappanl}\\
	\delta\omega_0 & = & \varDelta_\mathrm{p} - \sqrt{\left(\varDelta_\mathrm{p} - \mathcal{K}n_\mathrm{c} \right)\left(\varDelta_\mathrm{p} - 3\mathcal{K}n_\mathrm{c} \right) - \frac{\kappa_\mathrm{nl}^2 n_\mathrm{c}^2}{4}}.
	\label{eqn:Shift}
\end{eqnarray}
Subsequently, we use the $\delta\omega_0$ and $\kappa_\mathrm{p}$ as obtained from the measurements to determine the intracavity photon number $n_\mathrm{c}$ for each $P_\mathrm{p}$ without any knowledge of $\mathcal{K}$, cf. Supplementary Note~VI.
In panels (b) and (c) of Fig.~\ref{fig:Figure4}, the extracted values $\delta\omega_0$ and $\kappa_\mathrm{p}$ are shown for various bias flux values and plotted vs the intracircuit pump photon number $n_\mathrm{c}$ in device 3D$_1$ at a sample temperature $T_\mathrm{s} = 2.5\,$K.
Both quantities show a nearly linear dependence on pump photon number as implemented by our model and the corresponding slope depends in turn on the bias flux $\Phi_\mathrm{ext}$.
We fit the data using Eqs.~(\ref{eqn:Kappanl}, \ref{eqn:Shift}) and obtain $\mathcal{K}(\Phi_\mathrm{ext})$ for all three devices.
For all circuits, the nonlinearity shown in Fig.~\ref{fig:Figure4}(d) increases with increasing flux, but the absolute values differ by several orders of magnitude.
While the 2D circuit has a Kerr constant of only $\sim 1\,$Hz, the 3D circuits possess nonlinearities of $10^2$ to $10^3\,$Hz in 3D$_1$ and up to $10^3$ to $10^4\,$Hz in 3D$_2$, respectively.
All nonlinearities can still be considered small though and are in particular several orders of magnitude smaller than the cavity linewidths $\mathcal{K} \ll \kappa_0$.
We observe also very good agreement with the theoretical expression for the Kerr nonlinearity
\begin{equation}
	\mathcal{K} = -\frac{e^2}{2\hbar C_\mathrm{tot}}\left( \frac{L_\mathrm{J}}{2L + L_\mathrm{lin} + L_\mathrm{J}} \right)^3\left[ 1 + 3\Lambda\tan^2{\left(\pi\frac{\Phi}{\Phi_0}\right)} \right].
	\label{eqn:Kerr}
\end{equation}
with the electron charge $e$, the total circuit capacitance $C_\mathrm{tot} = C + C_\mathrm{c}$, the reduced Planck constant $\hbar$ and $\Lambda = (L_\mathrm{lin} + L_\mathrm{loop}/2)/(L_\mathrm{lin} + L_\mathrm{loop}/2 + L_\mathrm{J})$.
Note that we apply the method of nonlinear current conservation discussed in Ref.~\cite{Frattini18} to obtain this theoretical expression, cf. also Supplementary Note~X.
The unusual $\tan^2$ term in the last parenthesis of Eq.~(\ref{eqn:Kerr}) is however, not stemming from an asymmetry or a hidden third order nonlinearity, it is a correction factor for perfectly symmetric SQUIDs with screening parameter $\beta_\mathrm{L} > 0$.
How necessary it is to consider this extra term is revealed by the difference to the simple participation ratio expression (Eq.~(\ref{eqn:Kerr}) with $\Lambda = 0$), which is also shown in Fig.~\ref{fig:Figure4}(d) as dotted lines and which for large flux bias values differs from the exact result and the data by up to a factor of $\sim4$.

\subsection*{Discussion}
\vspace{-2mm}

In summary, we have reported and analyzed niobium-based superconducting quantum interference microwave circuits with integrated, monolithically fabricated 2D and 3D nano-constriction SQUIDs.
Our results revealed that the tuning range and the flux responsivity of the circuits increases with decreasing constriction thickness and critical current.
We have also characterized the Kerr anharmonicity of all circuits and found values between $0.5\,$Hz for the 2D circuits up to $10\,$kHz for the 3D circuit with the lowest critical current junctions.
The overall characteristics of the circuits make them highly promising candidates for quantum circuit and quantum sensing applications in particular when high dynamic range and high magnetic fields will be important such as in spin-qubit circuit quantum electrodynamics, hybrid quantum devices with magnonic oscillators or flux-mediated optomechanics.
The most interesting open questions to be investigated in future experiments are the circuit characteristics at temperatures in the mK regime and in large magnetic fields, the exact origin of the microwave losses in the nano-constriction circuits and possibilities to reduce them, as well as a theoretical and experimental investigation of the noise characteristics in such devices.
Finally, it will be interesting to investigate possibilities to further reduce the critical currents and the screening parameters, potentially by further reducing the size of the nano-constrictions in all three dimensions.

\subsection*{Acknowledgements}
\vspace{-2mm}
We gratefully acknowledge technical support by M. Turad and R. Löffler (both LISA$^+$) and by C. Back.
We acknowledge funding by the Deutsche Forschungsgemeinschaft (DFG) via grant numbers BO 6068/1-1, BO 6068/2-1 and KO 1303/13-2.

\subsection*{Data and code availability}
\vspace{-2mm}
All data and processing scripts presented in this paper will be available via Zenodo upon publication of this work.
Supplementary Material including the additional Refs.~\cite{Klein90, Igreja04, Wenner11, Rieger22, Rodrigues21} is appended.

\subsection*{Competing interest}
\vspace{-2mm}
The authors declare no competing interests.

\let\addcontentsline\oldaddcontentsline
\clearpage

\widetext
\begin{center}
	\noindent\textbf{\large Supplementary Material for:\\ Niobium Quantum Interference Microwave Circuits \\ with Monolithic Three-Dimensional (3D) Nanobridge Junctions}
	
	\normalsize
	\vspace{.3cm}
	
	\noindent{K.~Uhl \textit{et al.}}
	\\
\end{center}
\vspace{0.2cm}

\tableofcontents

\newpage
\setcounter{figure}{0}
\setcounter{equation}{0}
\setcounter{table}{0}
\renewcommand{\figurename}{Supplementary Figure}
\renewcommand{\tablename}{Supplementary Table}
\renewcommand{\theequation}{S\arabic{equation}}
\renewcommand{\bibnumfmt}[1]{[S#1]}
\renewcommand{\citenumfont}[1]{S#1}

\section{Supplementary Note I: Device fabrication}
\label{sec:Note1}
\vspace{-2mm}
In this note, the sample fabrication is described step-by-step including a schematic representation of the nano-constriction fabrication in Supplementary Fig.~\ref{fig:FigureS1}.  
\begin{figure*}[h]
	\centerline{\includegraphics[width=0.85\textwidth]{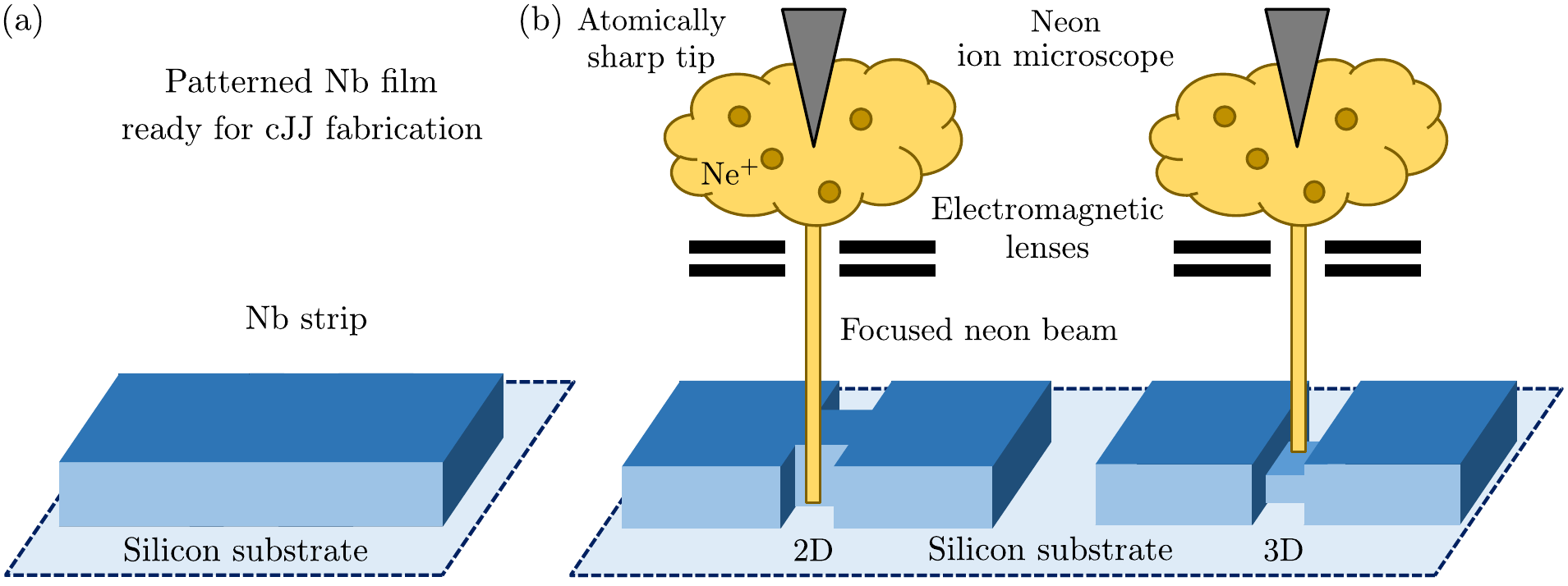}}
	\caption{\textsf{\textbf{Schematic niobium nano-constriction fabrication using a focused neon ion beam.} (a) shows a niobium (Nb) strip on a silicon substrate prepared for cutting a nano-constriction into it. (b) schematically shows the fabrication of the two different constriction types discussed in this study using a neon ion microscope (NIM); left 2D, right 3D constriction. The NIM consists of an atomically sharp tip maintained at a high voltage of $20\,$kV. Neon gas around the tip gets ionized and the ions are accelerated and focused through several electromagnetic lenses to form a beam with a nano-scaled spot size ($\sim 0.5\,$nm). For the 2D constrictions the strip is patterned by cutting two narrow slots from both sides into the strip. Therefore, the nanobridge has the same height as the remaining niobium leads. For the 3D constrictions the constriction also gets milled and thinned down from the top simultaneously with the slot cutting, just with a smaller ion dose.}}
	\label{fig:FigureS1}
\end{figure*}
\begin{itemize}
	\item \textbf{Step 1: Microwave cavity patterning.}
	\\
	The fabrication starts with sputtering $90\,$nm niobium (Nb) on top of a high-resistivity ($\rho > 10\,$k$\Omega\cdot\,$m) intrinsic two inch silicon wafer.
	The wafer thickness is $500\,$\textmu m.
	Then the complete wafer is covered with a $600\,$nm thick layer of ma-P 1205 photoresist by spin-coating and structured by means of maskless scanning laser photolithography ($\lambda_\mathrm{Laser} = 365\,$nm).
	After development of the resist in maD331/s for $40\,$s, the patterned side of the wafer gets etched by means of reactive ion etching using $\mathrm{SF}_6$.
	For cleaning, the wafer gets finally rinsed in multiple subsequent baths of acetone and isopropanol.
	\item \textbf{Step 2: Dicing and mounting for pre-characterization.}
	\\
	At the end of the microwave cavity fabrication, the wafer gets diced into individual $10\times10\,$mm$^2$ chips for mounting into a printed circuit board (PCB), where it gets wire-bonded to microwave feedlines and ground, and packaged into a radiation tight copper housing.
	After mounting into the measurement setup, the pre-characterization of the device is performed.
	\item \textbf{Step 3: 2D and 3D constriction fabrication.}
	\\
	Each pre-characterized LC circuit contains a square-shaped niobium loop structure with inner dimensions of $6\times6\,$\textmu m$^2$ and a conductor-strip-width of $3\,$\textmu m, cf. main paper Fig.~1.
	Two nano-constrictions are patterned into the opposite sides of the loops using the focused ion beam of a neon ion microscope (NIM).
	The NIM allows high-precision milling with a nano-scaled spot-size focused neon beam.   
	For the 2D constrictions, two $\sim20\,$nm narrow slot-shaped rectangles are ion-milled simultaneously from both sides into each loop strip with a dose of $20000\,$ions/nm$^2$ and an accelerating voltage of $20\,$kV.
	Simultaneous here means that the neon beam is scanned in a pattern that alternates repeatedly between the left and the right slot.
	In between the two slots, a milling gap of 40$\,$nm is left untouched: the later 2D constriction. 
	The 3D constriction patterning is performed in the same way, but additionally and again simultaneously to the slot rectangles the constriction is milled from the top with a third rectangle, but with a lower dose.
	The exact value of the constriction dose defines its remaining thickness.
	For the samples 3D$_1$ and 3D$_2$ of our work, the doses were $7500\,$ions/nm$^2$ and $8500\,$ions/nm$^2$, respectively.  
	\item \textbf{Step 4: Device mounting.}
	\\
	After the NIM cutting process the sample is mounted in the same way as in Step 2. 
\end{itemize}

\section{Supplementary Note II: Measurement setup}
\label{sec:Note2}
\begin{figure*}
	\centerline{\includegraphics[width=0.75\textwidth]{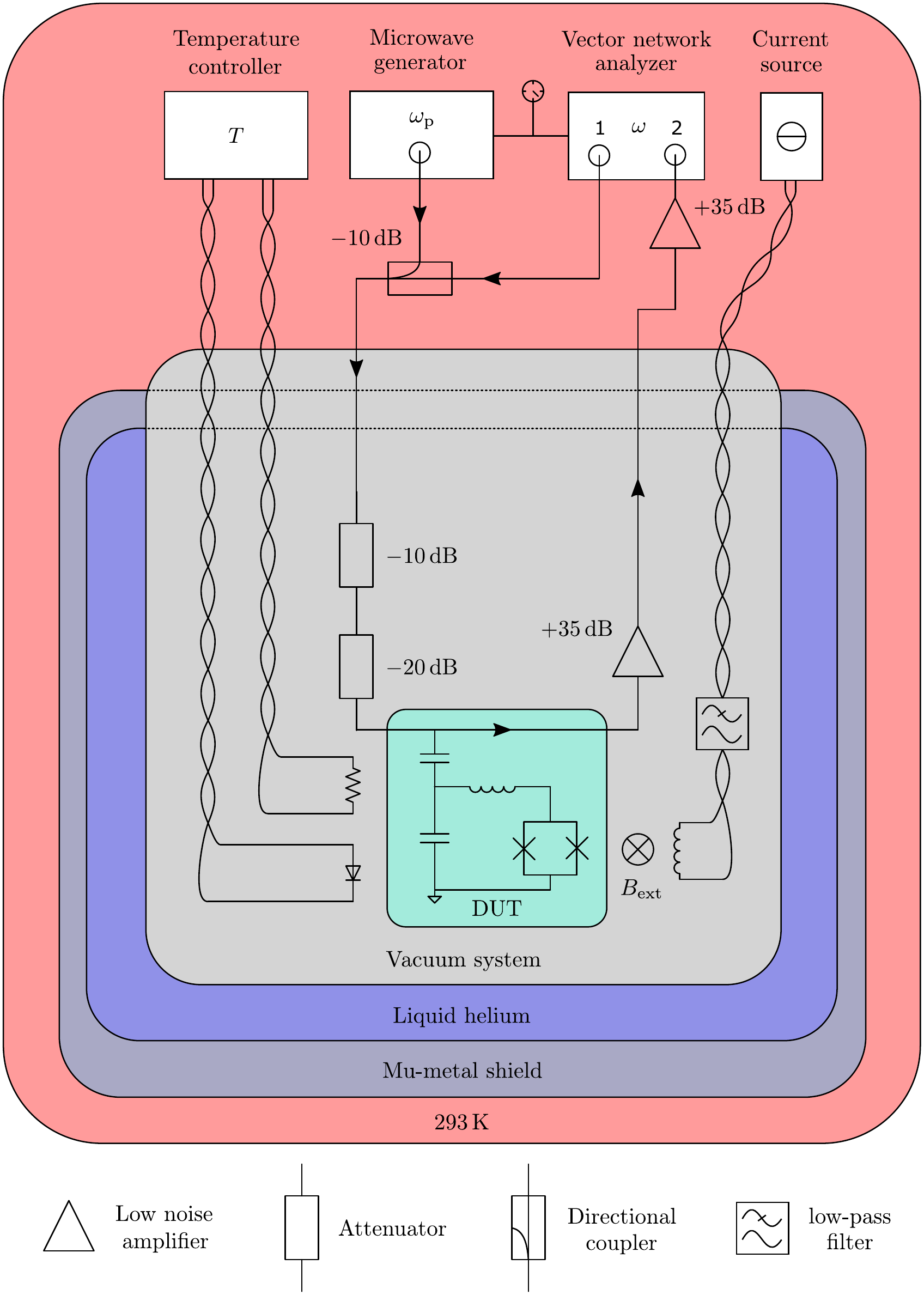}}
	\caption{\textsf{\textbf{Schematic of the measurement setup.} The isolation vacuum shield in between the helium container and the outer world/mu-metal shield is omitted. Detailed information is given in the text.}}
	\label{fig:FigureS2}
\end{figure*}
Both the junction-less circuits and the SQUID cavities, here generically labeled as device under test (DUT), are characterized in an evacuated sample space located at the end of a cryostat dipstick, which is introduced into a liquid helium cryostat.
The cryostat is covered by a double-layer room-temperature mu-metal shield to provide magnetic shielding for the whole sample space. 
A schematic illustration of the measurement setup is shown in Supplementary Fig.~\ref{fig:FigureS2}.
The DUT inside the copper housing is attached to a copper mounting bracket and a magnet coil for the application of a magnetic field perpendicular to the chip surface $B_\mathrm{ext}$.
The magnet coil is connected to a low-noise current source at room temperature with a twisted pair of copper wires.
The magnet wires get low-pass filtered at cryogenic temperature with a cutoff frequency at $\sim 3\,$kHz.  
Additionally the DUT is connected to two coaxial lines for input and output of the microwave signals.
A temperature diode is attached to the sample housing/the mounting bracket in close proximity to the actual sample and both are coupled to the liquid helium bath via the copper mounting bracket and through a small amount of helium exchange gas. 
For controlling the sample temperature $T_\mathrm{s}$, the diode is glued with varnish to the DUT copper housing and a manganine heating resistor (made of a twisted pair wire to avoid stray magnetic fields) is placed nearby. 
Both the temperature diode (4 wires) and the heating resistor (2 wires) are also connected via twisted pairs of copper wire to a temperature controller. 
The SQUID cavities are designed in a side-coupled geometry. 
Therefore the input and output signals were sent and received through two separate coaxial lines in order to measure the transmission spectrum $S_\mathrm{21}$ of the DUT by a vector network analyzer (VNA).  
The input line is attenuated by $-30\,$dB in order to balance the thermal radiation from room temperature to the cryostat temperature.
The attenuators are mounted in close proximity to the sample in the sample vacuum space and we assume them to have a temperature $T_\mathrm{att} \approx T_\mathrm{s}$.
For amplification of the weak microwave signal used here, a cryogenic high electron mobility transistor (HEMT) amplifier and a room temperature amplifier are mounted to the output line. 
The cryogenic HEMT is placed close to the DUT in order to minimize signal losses in between the sample and the amplifier chain.
For the two-tone experiment an additional fixed-frequency microwave pump tone with frequency $\omega_\mathrm{p}$ and power $P_\mathrm{p}$ is sent to the DUT.
This pump tone is generated by a microwave generator and combined via a $10\,$dB directional coupler with the VNA signal before entering the cryostat. 
In the experiment the VNA and microwave generator are both referenced to the $10\,$MHz clock of the generator. 
For cooling the device to temperatures below that of liquid helium $T_\mathrm{He} = 4.2\,$K, we pump at the helium dewar of the cryostat and reach down to $T_\mathrm{s, min} \lesssim 2.4\,$K with the current setup.
To achieve high-stability ($\Delta T_\mathrm{s} < 1 \,$mK) temperature control in the most relevant range for this work $2.4\,$K$\, \lesssim T_\mathrm{s} \lesssim 3.5\,$K, we use the helium pumping but additionally heat the sample with the heating resistor whose power is controlled via a PID feedback loop by the temperature controller.

\section{Supplementary Note III: Power calibration}
\label{sec:Note3}
\begin{figure*}[h]
	\centerline{\includegraphics[width=0.45\textwidth]{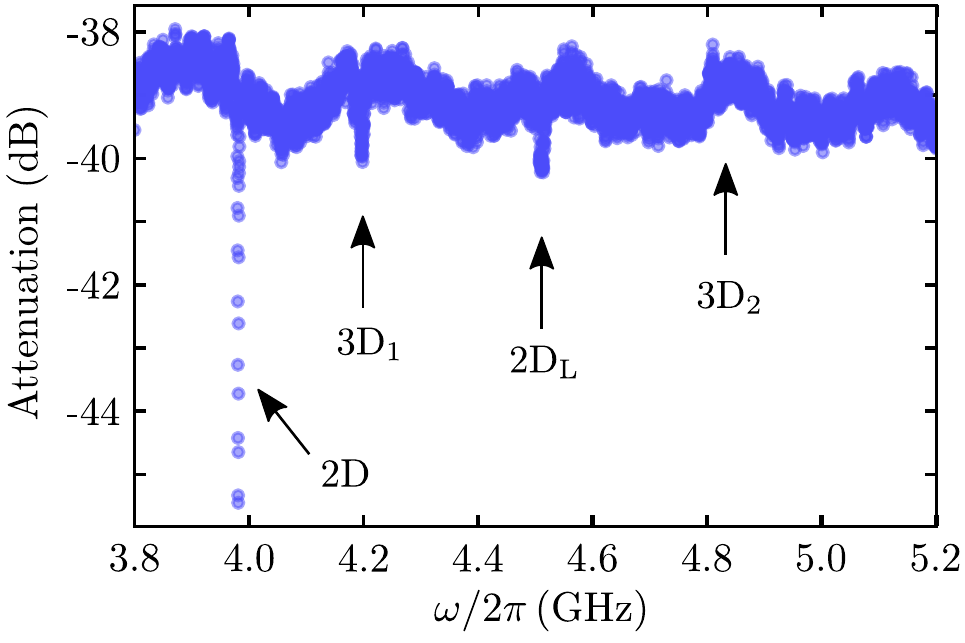}}
	\caption{\textsf{\textbf{Estimation of the frequency-dependent input line attenuation.} The attenuation is obtained by measuring 200 independent traces in the shown frequency range using the VNA and at $T_\mathrm{s} = 2.5\,$K. For each frequency point, the signal-to-noise ratio is determined as the mean-to-standard-deviation-ratio out of the 200 traces. In combination with the frequency-dependent HEMT noise temperature, the sample temperature and an estimated 1$\,$dB loss between the sample and the HEMT, we obtain the shown $\sim -39 \pm1\,$dB of attenuation between the VNA output and the sample. The frequency range contains the three SQUID cavities 2D, 3D$_1$ and 3D$_2$ presented in the main manuscript and an additional cavity 2D$_\mathrm{L}$ not discussed here. The resonance of 3D$_2$ is not visible without raw data processing here, cf. also Supplementary Note~\ref{sec:Note7}. }}
	\label{fig:FigureS3}
\end{figure*}
We use the input noise of the cryogenic HEMT amplifier in combination with the knowledge of the sample temperature (more precisely the temperature of the $20\,$dB attenuator, which is the one closer to the sample) as calibration method for the attenuation between the VNA output and the superconducting chip.
This allows to estimate the resulting input power on the on-chip feedline based on the set VNA output power. 
For each frequency point in the relevant frequency range $ \frac{\Delta\omega_\mathrm{att}}{2\pi} = 3.8-5.2\,$GHz the signal-to-noise ratio is determined from finding the mean and the standard deviation of 200 VNA traces.  
The total thermal noise power referenced to the input of the HEMT is given by
\begin{equation}
	P_\mathrm{HEMT} = 10 \log\left(\frac{k_\mathrm{B}\left[T_\mathrm{HEMT} + T_\mathrm{s}\right]}{1\,\mathrm{mW}}\right) + 10 \log\left(\frac{f_\mathrm{IFBW}}{\mathrm{Hz}}\right)
\end{equation}
where $k_\mathrm{B}$ is the Boltzmann constant, $T_\mathrm{HEMT}(\omega) = 7.46\,\mathrm{K} - \frac{3}{7\pi}\frac{\omega}{\mathrm{GHz}}\,\mathrm{K}$ is the noise temperature of the amplifier according to the specification datasheet in the frequency window $\Delta \omega_\mathrm{att}$ and $f_\mathrm{IFBW}$ is the measurement intermediate frequency bandwidth (IFBW).
We also estimate $1\,$dB loss between sample and HEMT.
In combination, we get the frequency-dependent input line attenuation as shown in Supplementary Fig.~\ref{fig:FigureS3}.
Note that the frequency range $\Delta \omega_\mathrm{att}$ contains four resonance dips, three belonging to the SQUID cavities 2D, 3D$_1$ and 3D$_2$ presented in the main manuscript and one additional SQUID cavity 2D$_\mathrm{L}$ with long 2D nano-constrictions that is not discussed here ($\omega_\mathrm{0} \approx 2\pi\cdot4.52\,$GHz). 
The resonance of 3D$_2$ is not easily visible without a raw data processing, cf. Supplementary Note~\ref{sec:Note7}. 
\section{Supplementary Note IV: The circuits without constrictions}
\label{sec:Note4}
\subsection{The lumped element circuit equivalent without the constrictions}
In the main manuscript, we discuss data from three distinct microwave circuits, one with 2D nanobridges and two devices with 3D nanobridges.
The basic design of the circuits comprises two interdigitated capacitors with multiple linear inductors, that are combined into a single one in our model for the sake of simplicity.
The interdigitated capacitors in all circuits have a finger length of $250\,$\textmu m, a finger and gap width of $3\,$\textmu m.
Only the number of fingers $N_\mathrm{idc}$ differs between the circuits.
Note that the finger number $N_\mathrm{idc}$ here refers to the total number of fingers in an IDC, not to the fingers per electrode.
We also remark that such a circuit has actually more than one mode, but we are only focusing on the fundamental mode here, i.e., the one with the lowest resonance frequency.
We model the circuit (before the constriction cutting) by a parallel RLC circuit with the inductance $L$, the capacitance $C$ and the resistance $R$.
Note that the inductance has both a geometric contribution $L_\mathrm{g}$ and a kinetic contribution $L_\mathrm{k}$ with $L = L_\mathrm{g} + L_\mathrm{k}$, as it is common for superconducting thin film circuits.
The parallel RLC circuit is coupled via a coupling capacitance $C_\mathrm{c}$ to the microwave feedline, which is a coplanar waveguide with characteristic impedance $Z_0 \approx 50\,\Omega$.
The total inductance $L$ contains also the contribution from the SQUID loop self inductance $L_\mathrm{loop}$ and so the circuit inductance without the loop would be given by $L - L_\mathrm{loop}/4$, for a schematic of the circuit (including the constriction inductances $L_\mathrm{c}$) see main paper Fig.~1(b).
The circuit resonance frequency before introducing the constrictions is given by
\begin{equation}
	\omega_\mathrm{b} = \frac{1}{\sqrt{L(C + C_\mathrm{c})}}
\end{equation}
and the internal and external linewidths are
\begin{equation}
	\kappa_\mathrm{i, b} = \frac{1}{R(C + C_\mathrm{c})}, ~~~~~ \kappa_\mathrm{e, b} = \frac{\omega_\mathrm{b}^2 C_\mathrm{c}^2 Z_0}{2(C + C_\mathrm{c})}
	\label{eqn:kappas}
\end{equation}
which are related to the corresponding quality factors via $Q_\mathrm{i, b} = \omega_\mathrm{b}/\kappa_\mathrm{i, b}$ and $Q_\mathrm{e, b} = \omega_\mathrm{b}/\kappa_\mathrm{e, b}$.
\subsection{Circuit parameters from simulations and measurements}
\label{sec:circuit_parameters}
\begin{figure*}
	\centerline{\includegraphics[width=0.85\textwidth]{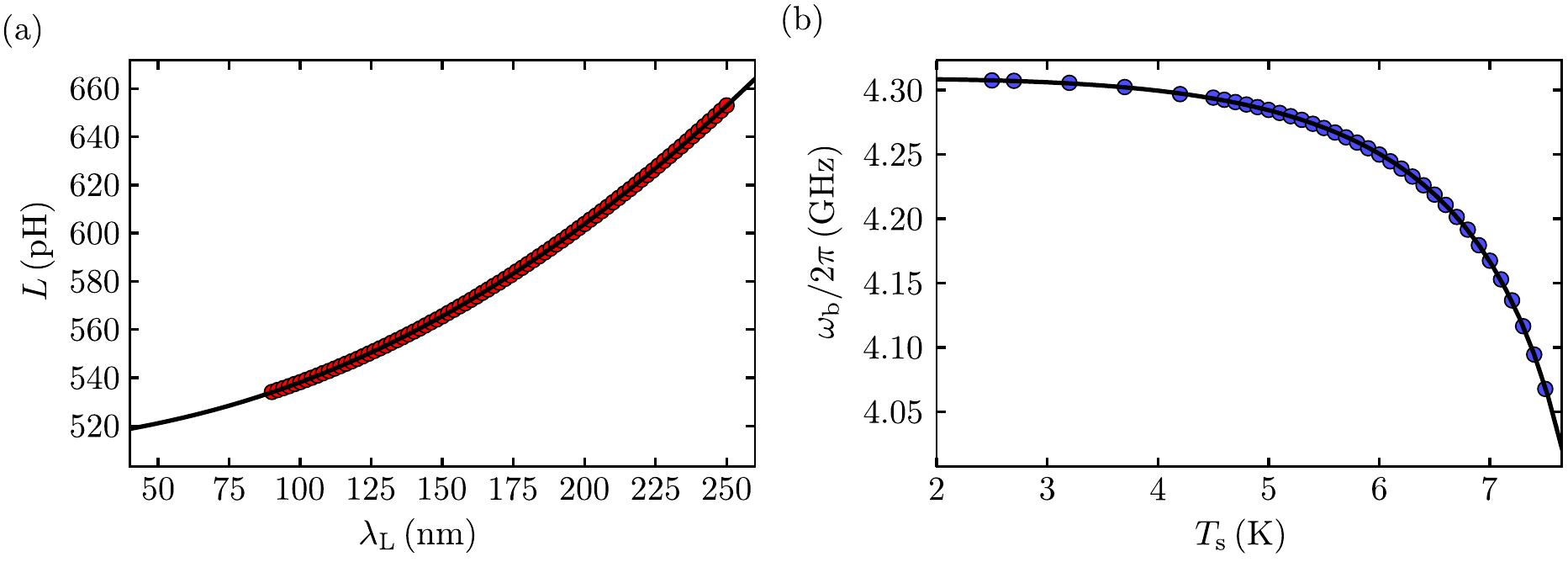}}
	\caption{\textsf{\textbf{Temperature dependence before junction cutting.} In (a) we show the total inductance $L(\lambda_\mathrm{L})$ of the cavity vs the London penetration depth $\lambda_\mathrm{L}$. The red circles are data obtained from 3D-MLSI simulations using different $\lambda_\mathrm{L}$ and the line is a fit using Eq.~(\ref{eqn:l_lambda}) with $L_\mathrm{g} = 511\,$pH and $g = 156$ as fit parameter. In panel (b), the cavity resonance frequency $\omega_\mathrm{b}$ is shown vs sample temperature $T_\mathrm{s}$. The shift in frequency with increasing temperature occurs due to a change of the total circuit inductance $L(T_\mathrm{s}) = L_\mathrm{g} + L_\mathrm{k}(T_\mathrm{s})$. Circles are data, line is a fit using Eq.~(\ref{eqn:omega_t}) and with $T_\mathrm{c} = 8.6\,$K, $C_\mathrm{tot} = 2.404\,$pF and $\lambda_0 = 153\,$nm as fit parameters. All data correspond to the circuit 3D$_1$.}}
	\label{fig:FigureS4}
\end{figure*}

The parameters we need for our circuits are the capacitances $C$ and $C_\mathrm{c}$, which in sum give $C_\mathrm{tot} = C + C_\mathrm{c}$.
On the inductive side, we need the total inductance $L$ and the loop inductance $L_\mathrm{loop}$.
The inductances, however, are both the sum of a geometric contribution and a kinetic contribution and for the case of the total inductance we get $L = L_\mathrm{g} + L_\mathrm{k}$ (analogously of course for the loop inductance).
What makes things even more complicated on one hand but also experimentally accessible on the other hand is that the kinetic contributions are dependent on the niobium London penetration depth $\lambda_\mathrm{L}$, which is a function of sample temperature $T_\mathrm{s}$.
We start our parameter extraction procedure by calculating the total linear inductance $L(\lambda_\mathrm{L}) = L_\mathrm{g} + L_\mathrm{k}(\lambda_\mathrm{L})$ with the software package 3D-MLSI \cite{Khapaev01_SI} for the range $90\,$nm$\,\leq \lambda_\mathrm{L} \leq 250\,$nm as this is typically the regime of the penetration depth of our sputtered $d_\mathrm{Nb} = 90\,$nm thick niobium films.
The relation between the penetration depth and the kinetic inductance is given by \cite{Klein90_SI}
\begin{equation}
	L_\mathrm{k} = \mu_0 g \lambda_\mathrm{eff}
\end{equation}
where $\mu_0$ is the vacuum permittivity, $g$ is a geometrical dimensionless factor taking into account the details of the superconducting structures and the effective penetration depth $\lambda_\mathrm{eff}$ of a thin film $d_\mathrm{Nb}\lesssim \lambda_\mathrm{L}$ is \cite{Klein90_SI}
\begin{equation}
	\lambda_\mathrm{eff} = \lambda_\mathrm{L}\coth{\frac{d_\mathrm{Nb}}{\lambda_\mathrm{L}}}.
\end{equation}
Then the total inductance is
\begin{equation}
	L = L_\mathrm{g} + \mu_0 g \lambda_\mathrm{L}\coth{\frac{d_\mathrm{Nb}}{\lambda_\mathrm{L}}}
	\label{eqn:l_lambda}
\end{equation}

which we use to fit the numerically obtained $L(\lambda_\mathrm{L})$ with $g$ and $L_\mathrm{g}$ as fit parameters, cf. Supplementary Fig.~\ref{fig:FigureS4}(a). 

In the experiment, we do not vary directly $\lambda_\mathrm{L}$ but the sample temperatur $T_\mathrm{s}$, the relation between the two is given by
\begin{equation}
	\lambda_\mathrm{L}(T_\mathrm{s}) = \frac{\lambda_0}{\sqrt{1 - \left(\frac{T_\mathrm{s}}{T_\mathrm{c}}\right)^4}}
\end{equation}
where $\lambda_0$ is the zero-temperature penetration depth and $T_\mathrm{c}$ is the critical temperature.
We measure the cavity resonance frequency $\omega_\mathrm{b}(T_\mathrm{s})$ and hence fit the experimentally obtained data with
\begin{eqnarray}
	\omega_\mathrm{b}(T_\mathrm{s}) & = & \frac{1}{\sqrt{C_\mathrm{tot}L(T_\mathrm{s})}} \nonumber \\
	& = & \frac{1}{\sqrt{C_\mathrm{tot}\left[ L_\mathrm{g} + \mu_0 g \frac{\lambda_0}{\sqrt{1 - \left(\frac{T_\mathrm{s}}{T_\mathrm{c}} \right)^4}}\coth{\left[ \frac{d_\mathrm{Nb}}{\lambda_0}\sqrt{1 - \left(\frac{T_\mathrm{s}}{T_\mathrm{c}}\right)^4} \right]} \right]}}
	\label{eqn:omega_t}
\end{eqnarray}
with $L_\mathrm{g}$ and $g$ as fixed parameters obtained as described above, $\mu_0 = 4\pi\cdot 10^{-7}\,$N$\cdot$A$^{-2}$ and $d_\mathrm{Nb} = 90\,$nm being constants and $C_\mathrm{tot}, \lambda_0$ and $T_\mathrm{c}$ as fit parameters.
Once we know $\lambda_0$, we can also calculate the loop inductance
\begin{equation}
	L_\mathrm{loop}(T_\mathrm{s}) = L_\mathrm{loop, g} + \mu_0 g_\mathrm{loop}\lambda_\mathrm{eff}(T_\mathrm{s})
\end{equation}
for all measurement temperatures.
Using 3D-MLSI the same way as for $L$ of the cavity, we obtain $L_\mathrm{loop,g} = 12.8\,$pH and $g_\mathrm{loop} = 12$.
For $T_\mathrm{s} = 2.5\,$K this means $L_\mathrm{loop} = 17\,$pH.
The next relevant parameter is the coupling capacitance $C_\mathrm{c}$, which we obtain from the measurement of the external linewidth $\kappa_\mathrm{e, b}$, the knowledge of $Z_0 \approx 50\,\Omega$, $\omega_\mathrm{b}$ and $C_\mathrm{tot} = C + C_\mathrm{c}$ by using Eq.~(\ref{eqn:kappas}).
Finally, we can also obtain $C = C_\mathrm{tot} - C_\mathrm{c}$.
To cross-check the values we obtain by this procedure, we also calculate the capacitances (I) by using the analytical expression derived in Ref.~\cite{Igreja04_SI} as well as (II) by doing Comsol Multiphysics simulations.
We find moderate deviations of $< 10\%$ between the experimental values (based on the inductance simulation) and the analytical expression and of $< 20\%$ between the experimental values and the Comsol simulations.
Since both, the analytical and the Comsol results overestimate the capacitance we obtain by the procedure described here, we think that over-etching into the Si substrate when doing the SF$_6$-etch of the Nb (especially in between the fingers), as well as a finger width slightly smaller than $3\,$\textmu m, can be identified as the source of these deviations.
Supplementary Table~\ref{tab:params} summarizes all relevant parameters of the three circuits, the temperature-dependent quantities $\omega_\mathrm{b}, \kappa_\mathrm{e, b}$ and $\kappa_\mathrm{i, b}$ are given at $T_\mathrm{s} = 2.5\,$K.

\begin{center}
	\begin{table}[h]
		\caption{\label{tab:params}\textsf{\textbf{Simulated and experimental parameters of the three circuits before cutting the nanobridge junctions}}. The finger number $N_\mathrm{idc}$ refers to the total number of fingers in each IDC. The geometric inductance $L_\mathrm{g}$ and the kinetic geometry factor $g$ are obtained from simulations with 3D-MLSI \cite{Khapaev01_SI}. From a fit to the temperature-dependence of $\omega_\mathrm{b}$ we obtain the zero-temperature penetration depth $\lambda_0$, the critical temperature $T_\mathrm{c}$ and the total capacitance $C_\mathrm{tot}$. From the measured external linwidth $\kappa_\mathrm{e, b}$ we subsequently find the coupling capacitance $C_\mathrm{c}$ and the circuit capacitance $C$. For completeness we also give $\kappa_\mathrm{i, b}$. All experimental values are given for $T_\mathrm{s} = 2.5\,$K. \\}
		\begin{tabular}{  l | l | l | l | l | l | l | l | l | l | l | l |}
			Resonator & $N_\mathrm{idc}$ & $L_\mathrm{g}\,$(pH) & $g$ & $\lambda_0\,$(nm) & $T_\mathrm{c}\,$(K) & $C_\mathrm{tot}\,$(pF) & $C\,$(pF) & $C_\mathrm{c}\,$(fF) & $\frac{\omega_\mathrm{b}}{2\pi}\,$ (GHz) &  $\frac{\kappa_\mathrm{e, b}}{2\pi}\,$(MHz) & $\frac{\kappa_\mathrm{i, b}}{2\pi}\,$(kHz) \\ \hline
			2D & 100 & 535 & 164 & 157 & 8.6 & 2.652 & 2.613 & 38 & 3.994 & 1.4 & 89 \\ \hline
			3D$_1$ & 92 & 511 & 156 & 153 & 8.6 & 2.404 & 2.373 & 31 & 4.308 & 1.2 & 73 \\ \hline
			3D$_2$ & 76 & 462 & 141 & 153 & 8.6 & 1.936 & 1.903 & 33 & 5.047 & 2.2 & 120 \\ \hline
		\end{tabular}
	\end{table}
\end{center}
In addition to the resonance frequency, we also extract the total resonance linewidth $\kappa_\mathrm{b}$ as a function of temperature, data for circuit 3D$_1$ are shown in Supplementary Fig.~\ref{fig:FigureS5}. 
At the elevated temperatures we are operating here $T_\mathrm{s} \gtrsim 0.3 T_\mathrm{c}$, the internal decay rate will be dominated by quasiparticle losses.
From the two-fluid model, the effective surface resistance of a superconductor with the corresponding correction factor for thin films and around the cavity resonance frequency is given by \cite{Klein90_SI}
\begin{equation}
	R_\mathrm{s, eff} = \frac{1}{2}\omega_\mathrm{b}^2 \mu_0^2 \lambda_\mathrm{L}^3 \sigma_\mathrm{n} \frac{n_\mathrm{n}}{n}\left[\coth{\left( \frac{d_\mathrm{Nb}}{\lambda_\mathrm{L}}\right)} + \frac{d_\mathrm{Nb}/\lambda_\mathrm{L}}{\sinh^2{\left( \frac{d_\mathrm{Nb}}{\lambda_\mathrm{L}}\right)}} \right],
\end{equation}
where $\sigma_n$ is the normal state conductivity, $n_\mathrm{n}$ is the quasiparticle density and $n = n_\mathrm{n} + n_\mathrm{s}$ is the total electron density with $n_\mathrm{s}$ being the superconducting charge carrier density (twice the Cooper pair density).
The temperature dependence of the quasiparticle density is given by
\begin{equation}
	\frac{n_\mathrm{n}(T_\mathrm{s})}{n} = \left(\frac{T_\mathrm{s}}{T_\mathrm{c}}\right)^4.
\end{equation}
Since the quasiparticle loss channel is equivalent to the inductive channel, the resulting circuit model lumped element resistance $R' \propto R_\mathrm{s, eff}$ is expected to be in series with $L$.
We can, however, transform this into the parallel resistor $R$ via
\begin{equation}
	R = \frac{R'^2 + \omega_\mathrm{b}^2 L^2}{R'} \approx \frac{\omega_\mathrm{b}^2 L^2}{R'}
\end{equation}
where we used $\omega_\mathrm{b}^2L^2 \gg R'^2$.
Combining this result with Eq.~(\ref{eqn:kappas}) we get
\begin{eqnarray}
	\kappa_\mathrm{i, b}(T_\mathrm{s}) & = & \omega_\mathrm{b}^2(T_\mathrm{s}) R'(T_\mathrm{s}) C_\mathrm{tot} \nonumber \\
	& = & A_\kappa\omega_\mathrm{b}^4(T_\mathrm{s})\lambda_\mathrm{L}^3(T_\mathrm{s})\left( \frac{T_\mathrm{s}}{T_\mathrm{c}}\right)^4\left[\coth{\left( \frac{d_\mathrm{Nb}}{\lambda_\mathrm{L}(T_\mathrm{s})}\right)} + \frac{d_\mathrm{Nb}/\lambda_\mathrm{L}(T_\mathrm{s})}{\sinh^2{\left( \frac{d_\mathrm{Nb}}{\lambda_\mathrm{L}(T_\mathrm{s})}\right)}} \right]
	\label{eqn:kappa_qp}
\end{eqnarray}
with the fit parameter $A_\kappa$ that contains geometry, material properties and other temperature-independent contributions.

\begin{figure*}
	\centerline{\includegraphics[width=0.40\textwidth]{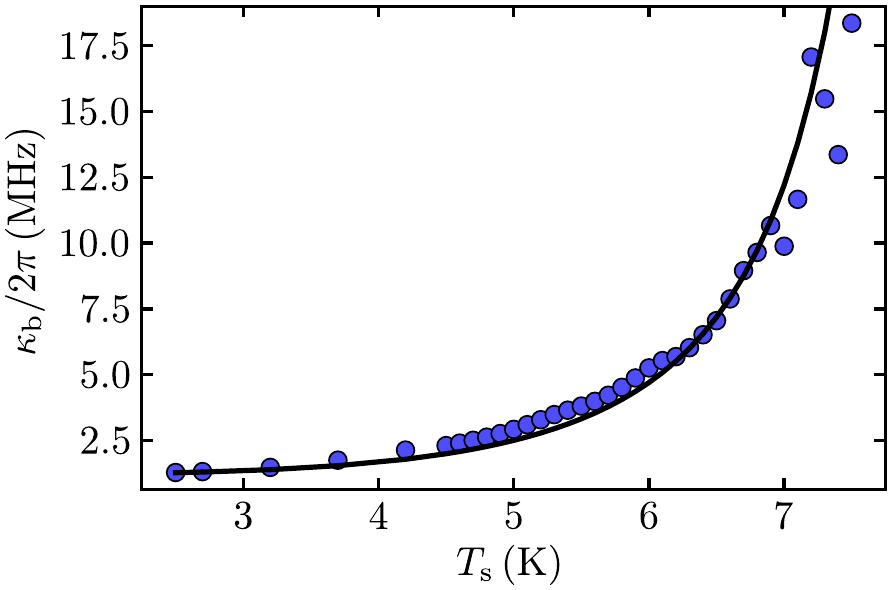}}
	\caption{\textsf{\textbf{Temperature dependence of linewidth indicates quasiparticle losses.} Total circuit linewidth $\kappa_\mathrm{b}/2\pi$ vs sample temperature $T_\mathrm{s}$ before cutting the nano-constrictions. The linewidth increases with increasing temperature, indicating growing losses by thermal quasiparticles in the superconductor. Circles are data for device 3D$_1$, line is a fit using Eqs.~(\ref{eqn:kappa_qp}, \ref{eqn:kappa_T}) with $A_\kappa$ as single fit parameter.}}
	\label{fig:FigureS5}
\end{figure*}

Since we are not certain that we can reliably discriminate between $\kappa_\mathrm{i, b}$ and $\kappa_\mathrm{e, b}$ due to cable resonances and impedance mismatches in the setup leading to Fano interferences, we fit the temperature dependence of the total linewidth using
\begin{equation}
	\kappa_\mathrm{b}(T_\mathrm{s}) = \kappa_\mathrm{e, b} + \kappa_\mathrm{i, b}(T_\mathrm{s})
	\label{eqn:kappa_T}
\end{equation}
with $\kappa_\mathrm{e, b}$ being a constant value we obtain for $T_\mathrm{s} = 2.5\,$K.
The theoretical variation of $\kappa_\mathrm{e, b}$ for large temperatures would only be $\sim 140\,$kHz and is therefore negligible compared to the total linewidth.
The agreement between the experimental data and the fit line is very good, cf. Supplementary Fig.~\ref{fig:FigureS5}, and we obtain nearly identical curves and agreemeents for all three resonators discussed in this work.

\section{Supplementary Note V: The circuits with constrictions}
\label{sec:Note5}
\subsection*{The lumped element circuit equivalent with the constrictions}
We observe that cutting the constrictions into the circuit leads to a shift of the resonance frequency and to a broadening of the resonance linewidth.
In agreement with the two-fluid model, we therefore model the circuit elements introduced by the junction as a constriction inductance $L_\mathrm{c}$ in parallel with a constriction resistance $R_\mathrm{c}$.
The inductance in turn we split into a series combination of a nonlinear Josephson part $L_\mathrm{J}$ and a linear contribution $L_\mathrm{lin}$, cf. Fig.~1(b) of the main paper.
This combination leads to a forward-skewed sinusoidal current-phase relation (CPR), where the skewedness is related to the ratio $L_\mathrm{J0}/L_\mathrm{lin}$ with $L_\mathrm{J0} = \frac{\Phi_0}{2\pi I_0}$, $\Phi_0 \approx 2.068\cdot 10^{-15}\,$Tm$^2$ and the critical constriction current $I_0$.
Such a skewed CPR is typical for constriction-type Josephson junctions (cJJ) \cite{Likharev79_SI, Hasselbach02_SI}.
We note that we omit any additional capacitance, as according to our simulations the impedance of a possible constriction capacitance is negligible compared to its inductance impedance.
Hence, the input impedance of a single constriction around the cavity resonance frequency $\omega \approx \omega_0$ is given by
\begin{equation}
	\frac{1}{Z_\mathrm{c}} = \frac{1}{R_\mathrm{c}} + \frac{1}{i\omega_0 L_\mathrm{c}},
\end{equation}
which leads to the total SQUID inductance of
\begin{equation}
	Z_\mathrm{S} = \frac{1}{2} \frac{i\omega_0 L_\mathrm{c}R_\mathrm{c}}{R_\mathrm{c} + i\omega_0 L_\mathrm{c}} + i\omega_0\frac{L_\mathrm{loop}}{4}.
\end{equation}
The originally purely inductive circuit branch then has the input impedance
\begin{equation}
	Z_L = i\omega_0 L + \frac{1}{2} \frac{i\omega_0 L_\mathrm{c}R_\mathrm{c}}{R_\mathrm{c} + i\omega_0 L_\mathrm{c}}.
\end{equation}
If we want to work with a low level of approximation, we transform this impedance into a series impedance of a single inductor $L_+$ and a resistor $R_+$ by
\begin{equation}
	R_+ + i\omega_0L_+ = \frac{1}{2}\frac{\omega_0^2 L_\mathrm{c}^2 R_\mathrm{c}}{R_\mathrm{c}^2 + \omega_0^2 L_\mathrm{c}^2} + i\omega_0\left(L + \frac{1}{2} L_\mathrm{c}\frac{R_\mathrm{c}^2}{R_\mathrm{c}^2 + \omega_0^2 L_\mathrm{c}^2} \right)
\end{equation}
As next step and in order to easily integrate this impedance branch into the formalism used here, we will now transform it into a parallel combination of an inductance $L^*$ and a resistance $R^*$
\begin{equation}
	\frac{1}{R^*} + \frac{1}{i\omega_0 L^*} = \frac{1}{R_+ + i\omega_0L_+}
\end{equation}
from which we find
\begin{equation}
	R^* = \frac{R_+^2 + \omega_0^2 L_+^2}{R_+}, ~~~~~ L^* = \frac{R_+^2 + \omega_0^2 L_+^2}{\omega_0^2L_+}.
\end{equation}
We approximate this now using $R_+^2 \ll \omega_0^2 L_+^2$ as
\begin{equation}
	R^* \approx \frac{\omega_0^2 L_+^2}{R_+}, ~~~~~ L^* \approx L_+
\end{equation}
The total linewidth is now given as
\begin{eqnarray}
	\kappa_\mathrm{i} & = & \frac{1}{C + C_\mathrm{c}}\frac{1}{R} +  \frac{1}{C + C_\mathrm{c}}\frac{1}{R^*} \\
	& = & \kappa_\mathrm{i, b} + \kappa_\mathrm{c} \\
	& \approx & \kappa_\mathrm{c}
\end{eqnarray}
and the resonance frequency as
\begin{equation}
	\omega_0 = \frac{1}{\sqrt{\left( C + C_\mathrm{c}\right)L^*}}
\end{equation}
From measuring $\omega_0$ and $\kappa_\mathrm{c}$ we can then determine $R_\mathrm{c}$ and $L_\mathrm{c}$.
For our manuscript, we can even work with a stronger approximation and still only have about $1\%-2\%$ error (obtained from comparing with exact results) in the extracted values, mainly for large flux biases $\Phi_\mathrm{ext}/\Phi_0 \sim \pm 0.5$.
Considering additionally $R_\mathrm{c}^2 \gg \omega_0^2 L_\mathrm{c}^2$ we find the expressions
\begin{equation}
	R^* \approx 2\frac{\left(L + \frac{L_\mathrm{c}}{2} \right)^2}{L_\mathrm{c}^2}R_\mathrm{c}, ~~~~~ L^* = L + \frac{L_\mathrm{c}}{2}.
\end{equation}
With these, we can write for the internal linewidth and the resonance frequency
\begin{equation}
	\kappa_\mathrm{i} = \frac{\omega_0^2 L_\mathrm{c}^2}{2L + L_\mathrm{c}}\frac{1}{R_\mathrm{c}}, ~~~~~ \omega_0 = \frac{1}{\sqrt{\left( C + C_\mathrm{c} \right)\left( L + \frac{L_\mathrm{c}}{2} \right)}}.
\end{equation}
\section{Supplementary Note VI: Circuit response model}
\label{sec:Note6}
\subsection{Equation of motion and general considerations}
We model the classical intracavity field $\alpha$ of the SQUID circuits with Kerr nonlinearity and nonlinear damping using the equation of motion
\begin{equation}
	\dot{\alpha} = \left[i(\omega_\mathrm{c} + \mathcal{K}|\alpha|^2) - \frac{\kappa + \kappa_\mathrm{nl}|\alpha|^2}{2}\right]\alpha + i\sqrt{\frac{\kappa_\mathrm{e}}{2}}S_\mathrm{in}.
\end{equation}
Here, $\omega_\mathrm{c}$ is the cavity resonance frequency ($=\omega_\mathrm{b}$ before cutting and $=\omega_0$ after), $\mathcal{K}$ is the Kerr nonlinearity (frequency shift per photon), $\kappa$ is the bare total linewidth ($=\kappa_\mathrm{b}$ before cutting and $=\kappa_0$ after), $\kappa_\mathrm{nl}$ is the nonlinear damping constant, $\kappa_\mathrm{e}$ is the external linewidth ($=\kappa_\mathrm{e, b}$ before cutting) and $S_\mathrm{in}$ is the input field.
The intracavity field is normalized such that $|\alpha|^2 = n_\mathrm{c}$ corresponds to the intracavity photon number $n_\mathrm{c}$ and $|S_\mathrm{in}|^2$ to the input photon flux (photons per second) on the coplanar waveguide feedline.
The solution of this equation of motion depends significantly on the pump power and the number of tones sent to the cavity now.
The ideal transmission response function, however, will always be of the form
\begin{equation}
	S_{21}^\mathrm{ideal} = 1 + i\sqrt{\frac{\kappa_\mathrm{e}}{2}}\frac{\alpha}{S_\mathrm{in}}
\end{equation}
with the solution of interest $\alpha$.

\subsection{The linear single-tone regime}

In the linear single-tone regime, we set $\mathcal{K} = \kappa_\mathrm{nl} = 0$.
Then, we can solve the remaining equation by Fourier transform and obtain
\begin{equation}
	\alpha = \frac{i\sqrt{\frac{\kappa_\mathrm{e}}{2}}}{\frac{\kappa}{2} + i(\omega - \omega_\mathrm{c})}S_\mathrm{in}
\end{equation}
The ideal transmission response of a capacitively side-coupled and linear LC circuit is then given by
\begin{equation}
	S_{21}^\mathrm{ideal} = 1 - \frac{\kappa_\mathrm{e}}{\kappa + 2i(\omega - \omega_\mathrm{c})}.
	\label{eqn:S21ideal}
\end{equation}

\subsection{The nonlinear single-tone regime}

In the nonlinear single-tone regime, we have to solve the full equation of motion and start by setting the input field to $S_\mathrm{in,st} = S_0e^{i\phi_\mathrm{p}}e^{i\omega t}$ with real-valued $S_0$.
For the intracavity field, we make the Ansatz $\alpha(t) = \alpha_0e^{i\omega t}$ with real-valued $\alpha_0$.
The phase delay between input and response is fully encoded in $\phi_\mathrm{p}$.
Then the equation of motion reads
\begin{equation}
	i\omega \alpha_0 = \left[i\left(\omega_\mathrm{c} + \mathcal{K}\alpha_0^2\right) - \frac{\kappa + \kappa_\mathrm{nl}\alpha_0^2}{2}\right]\alpha_0 + i\sqrt{\frac{\kappa_\mathrm{e}}{2}}S_0e^{i\phi_\mathrm{p}}
\end{equation}
which after multiplication with its complex conjugate yields the characteristic polynomial for the intracircuit photon number $n_\mathrm{c} = \alpha_0^2$
\begin{equation}
	n_\mathrm{c}^3\left[\mathcal{K}^2 + \frac{\kappa_\mathrm{nl}^2}{4}\right] + n_\mathrm{c}^2\left[ \frac{\kappa \kappa_\mathrm{nl}}{2} - 2\mathcal{K}\varDelta \right] + n_\mathrm{c}\left[\varDelta^2 + \frac{\kappa^2}{4}\right] - \frac{\kappa_\mathrm{e}}{2}S_0^2 = 0.
	\label{eqn:ploy_st}
\end{equation}
Here $\varDelta = \omega - \omega_\mathrm{c}$ is the detuning between the microwave input tone and the bare cavity resonance.
The real-valued roots of this polynomial correspond to the physical solutions for the amplitude $\alpha_0$, the highest and lowest amplitudes are the stable states in the case of three real-valued roots.
For the complete complex transmission, we also need the phase $\phi_\mathrm{p}$, which we obtain via
\begin{equation}
	\phi_\mathrm{p} = \atan2\left(-\frac{\kappa + \kappa_\mathrm{nl}n_\mathrm{c}}{2}, \varDelta - \mathcal{K}n_\mathrm{c}\right)
\end{equation}
Having both parts of the complex field solution at hand, we can also calculate the transmission

\begin{eqnarray}
	S_{21, \mathrm{nl}}^\mathrm{ideal} & = & 1 + i\sqrt{\frac{\kappa_\mathrm{e}}{2}}\frac{\alpha}{S_\mathrm{in,st}} \nonumber \\
	& = & 1 + i\sqrt{\frac{\kappa_\mathrm{e}}{2}}\frac{\alpha_0}{S_\mathrm{0}}e^{-i\phi_\mathrm{p}}.
\end{eqnarray}
Note that we do not use these equations for any data analysis in this manuscript, but we include them for paedagogical reasons, since they facilitate understanding the two-tone regime.

\subsection{The linearized two-tone regime}

In the two-tone experiments, we apply a strong pump tone with fixed frequency $\omega_\mathrm{p}$ and fixed power $P_\mathrm{p}$ and probe the cavity response with a weak additional scanning tone, the total input then is $S_\mathrm{in, tt} = S_0e^{i\phi_\mathrm{p}}e^{i\omega_\mathrm{p}t} + S_\mathrm{pr}(t)e^{i\omega_\mathrm{p} t}$.
The probe input amplitude $S_\mathrm{pr}(t)$ is time-dependent and complex-valued.
As Ansatz for the intracavity field we choose $\alpha(t) = \alpha_0e^{i\omega_\mathrm{p}t} + \alpha_\mathrm{pr}(t)e^{i\omega_\mathrm{p} t}$ with a complex and time-dependent $\alpha_\mathrm{pr}(t)$ and obtain the equation of motion
\begin{eqnarray}
	i\omega_\mathrm{p}\alpha_0 + i\omega_\mathrm{p}\alpha_\mathrm{pr} + \dot{\alpha}_\mathrm{pr} & = & i\left[\omega_\mathrm{c} + \mathcal{K}\left(\alpha_0^2 + \alpha_0(\alpha_\mathrm{pr} + \alpha_\mathrm{pr}^*) + |\alpha_\mathrm{pr}|^2\right)\right]\alpha_0 \nonumber \\
	& & + i\left[\omega_\mathrm{c} + \mathcal{K}\left(\alpha_0^2 + \alpha_0(\alpha_\mathrm{pr} + \alpha_\mathrm{pr}^*) + |\alpha_\mathrm{pr}|^2\right)\right]\alpha_\mathrm{pr} \nonumber \\
	& & -\left[\frac{\kappa}{2} + \frac{\kappa_\mathrm{nl}}{2}\left(\alpha_0^2 + \alpha_0(\alpha_\mathrm{pr} + \alpha_\mathrm{pr}^*) + |\alpha_\mathrm{pr}|^2\right)\right]\alpha_0 \nonumber \\
	& & -\left[\frac{\kappa}{2} + \frac{\kappa_\mathrm{nl}}{2}\left(\alpha_0^2 + \alpha_0(\alpha_\mathrm{pr} + \alpha_\mathrm{pr}^*) + |\alpha_\mathrm{pr}|^2\right)\right]\alpha_\mathrm{pr} \nonumber \\
	& & + i\sqrt{\frac{\kappa_\mathrm{e}}{2}}S_0e^{i\phi_\mathrm{p}} + i\sqrt{\frac{\kappa_\mathrm{e}}{2}}S_\mathrm{pr}.
\end{eqnarray}
Now we perform the linearization, i.e., we drop all terms not linear in the small quantity $\alpha_\mathrm{pr}$ and get 
\begin{eqnarray}
	i\omega_\mathrm{p}\alpha_0 + i\omega_\mathrm{p}\alpha_\mathrm{pr} + \dot{\alpha}_\mathrm{pr} & = & \left[i\left(\omega_\mathrm{c} + \mathcal{K}n_\mathrm{c} \right) - \frac{\kappa + \kappa_\mathrm{nl}n_\mathrm{c}}{2}\right]\alpha_0 \nonumber \\
	& & + \left[i\left(\omega_\mathrm{c} + 2\mathcal{K}n_\mathrm{c} \right) - \frac{\kappa + 2\kappa_\mathrm{nl}n_\mathrm{c}}{2}\right]\alpha_\mathrm{pr} \nonumber \\
	& & +\left[i\mathcal{K} - \frac{\kappa_\mathrm{nl}}{2}\right]n_\mathrm{c}\alpha_\mathrm{pr}^* \nonumber \\
	& & + i\sqrt{\frac{\kappa_\mathrm{e}}{2}}S_0e^{i\phi_\mathrm{p}} + i\sqrt{\frac{\kappa_\mathrm{e}}{2}}S_\mathrm{pr}.
\end{eqnarray}
The time-independent terms are identical to the Eq.~(\ref{eqn:ploy_st}) of the nonlinear single-tone experiment and allow to determine $\alpha_0$ and $n_\mathrm{c}$ via the characteristic polynomial now.
The remaining equation can be Fourier transformed to give
\begin{eqnarray}
	\frac{\alpha_\mathrm{pr}}{\chi_\mathrm{pr}} & = & \left[i\mathcal{K} - \frac{\kappa_\mathrm{nl}}{2}\right]n_\mathrm{c}\overline{\alpha}_\mathrm{pr} +  i\sqrt{\frac{\kappa_\mathrm{e}}{2}}S_\mathrm{pr}.
\end{eqnarray}
where
\begin{equation}
	\chi_\mathrm{pr} = \frac{1}{\frac{\kappa + 2\kappa_\mathrm{nl}n_\mathrm{c}}{2} + i\left(\varDelta_\mathrm{p} - 2\mathcal{K}n_\mathrm{c} + \varOmega  \right)}
\end{equation}
with $\overline{\alpha}_\mathrm{pr} = \alpha_\mathrm{pr}^*(-\varOmega)$, the detuning between pump and bare cavity resonance $\varDelta_\mathrm{p} = \omega_\mathrm{p} - \omega_\mathrm{c}$ and the probe frequency with respect to the pump $\varOmega = \omega - \omega_\mathrm{p}$.
Using the equivalent equation for $\overline{\alpha}_\mathrm{pr}$ with $\overline{S}_\mathrm{pr} = 0$, we get
\begin{eqnarray}
	\alpha_\mathrm{pr} & = &  i\chi_\mathrm{g}\sqrt{\frac{\kappa_\mathrm{e}}{2}}S_\mathrm{pr}.
\end{eqnarray}
with
\begin{equation}
	\chi_\mathrm{g} = \frac{\chi_\mathrm{pr}}{1 - \left[\mathcal{K}^2 + \frac{\kappa_\mathrm{nl}^2}{4}\right]n_\mathrm{c}^2\chi_\mathrm{pr}\overline{\chi}_\mathrm{pr}}
\end{equation}
and for the two-tone transmission parameter
\begin{equation}
	S_\mathrm{21, tt}^\mathrm{ideal} = 1 - \frac{\kappa_\mathrm{e}}{2}\chi_\mathrm{g}.
\end{equation}

\subsection{The pumped Kerr-modes}

To find the resonance frequencies of the susceptibility $\chi_\mathrm{g}$, we solve the complex frequency for which $\chi_\mathrm{g}^{-1} = 0$. The condition is 
\begin{equation}
	1 - \left[\mathcal{K}^2 + \frac{\kappa_\mathrm{nl}^2}{4}\right]n_\mathrm{c}^2\chi_\mathrm{pr}\overline{\chi}_\mathrm{pr} = 0
\end{equation}
which is solved by 
\begin{equation}
	\tilde{\omega}_\mathrm{1,2} = \omega_\mathrm{p} + i\frac{\kappa + 2 \kappa_\mathrm{nl} n_\mathrm{c}}{2} \pm  \sqrt{\left(\varDelta_\mathrm{p} - \mathcal{K} n_\mathrm{c}\right)\left(\varDelta_\mathrm{p} - 3 \mathcal{K} n_\mathrm{c}\right) - \frac{\kappa_\mathrm{nl}^2 n_\mathrm{c}^2}{4}}.
\end{equation}
The real part corresponds to the resonance frequency $\omega_\mathrm{1,2} = \Re\left(\tilde{\omega}_\mathrm{1,2}\right)$ and the imaginary part corresponds to half the mode linewidth $\kappa_\mathrm{1,2} = 2 \Im\left(\tilde{\omega}_\mathrm{1,2}\right)$. 
So in the regime where the argument of the square root is $>0$ (always true for our experimental parameters), the system has two resonances
\begin{equation}
	\omega_\mathrm{1,2} = \omega_\mathrm{p} \pm \sqrt{\left(\varDelta_\mathrm{p} - \mathcal{K} n_\mathrm{c}\right)\left(\varDelta_\mathrm{p} - 3 \mathcal{K} n_\mathrm{c}\right) - \frac{\kappa_\mathrm{nl}^2 n_\mathrm{c}^2}{4}},
\end{equation}
split symmetrically around the pump frequency.
In the experiment and with the parameters we are using, we only observe one of the two modes though, the one at $\omega_2 = \omega_0'$.
The shift of this mode with respect to its unpumped frequency $\omega_0$ is given by
\begin{equation}
	\delta\omega_0 = \varDelta_\mathrm{p} - \sqrt{\left(\varDelta_\mathrm{p} - \mathcal{K} n_\mathrm{c}\right)\left(\varDelta_\mathrm{p} - 3 \mathcal{K} n_\mathrm{c}\right) - \frac{\kappa_\mathrm{nl}^2 n_\mathrm{c}^2}{4}}.
	\label{eqn:ttshift}
\end{equation}
When we measure the pumped resonance, we also extract the pumped linewidth
\begin{equation}
	\kappa_\mathrm{p} = \kappa_0 + 2\kappa_\mathrm{nl}n_\mathrm{c}
\end{equation}
and so the only free parameter when fitting the resonance frequency shift vs pump photon number (see next subsection) using Eq.~(\ref{eqn:ttshift}) is the Kerr constant $\mathcal{K}$.
For brevity, we also introduce the short version
\begin{equation}
	\kappa_1 = \kappa_\mathrm{nl}n_\mathrm{c}.
\end{equation}
\subsection{The intracircuit pump photon number}
One might expect we need to know the value of $\mathcal{K}$ to calculate the intracircuit pump photon number from the pump-induced frequency shift and linewidth broadening.
This is not the case though, which allows us to first determine $n_\mathrm{c}$ and subsequently fit the frequency shift $\delta\omega_0$ as function of $n_\mathrm{c}$ to extract $\mathcal{K}$ from the data.
We start by setting $\mathcal{K}n_\mathrm{c} = x$ and then solve the characteristic polynomial  Eq.~(\ref{eqn:ploy_st}) for $x$.
We get (assuming $n_\mathrm{c}> 0$)
\begin{equation}
	x_\mathrm{1/2} = \varDelta_\mathrm{p} \pm \sqrt{\frac{\kappa_\mathrm{e}}{2}\frac{n_\mathrm{in}}{n_\mathrm{c}} - \frac{\kappa_\mathrm{eff}^2}{4}}
\end{equation}
where $n_\mathrm{in} = S_0^2$ and $\kappa_\mathrm{eff} = \kappa_0 + \kappa_1$.
The solution we are interested in is $x_2$.
Injecting this into Eq.~(\ref{eqn:ttshift}), the frequency relative to the drive
\begin{equation}
	\delta = \sqrt{\left( \varDelta_\mathrm{p} - x_2 \right)\left( \varDelta_\mathrm{p} - 3x_2\right) - \frac{\kappa_1^2}{4}}
\end{equation}
leads to
\begin{equation}
	\delta^2 = \sqrt{\frac{\kappa_\mathrm{e}}{2}\frac{n_\mathrm{in}}{n_\mathrm{c}} - \frac{\kappa_\mathrm{eff}^2}{4}}\left(3\sqrt{\frac{\kappa_\mathrm{e}}{2}\frac{n_\mathrm{in}}{n_\mathrm{c}} - \frac{\kappa_\mathrm{eff}^2}{4}} - 2\varDelta_\mathrm{p}  \right).
\end{equation}
We can solve this equation for $n_\mathrm{c}$ and find
\begin{equation}
	n_\mathrm{c} = \frac{2P_\mathrm{p}}{\hbar\omega_\mathrm{p}}\frac{\kappa_\mathrm{e}}{\kappa_\mathrm{eff}^2 + 4\tilde{\varDelta}^2}
	\label{eqn:nc_nl}
\end{equation}
with the effective detuning
\begin{equation}
	\tilde{\varDelta}^2 = \frac{2}{9}\left[\varDelta_\mathrm{p}^2 + \varDelta_\mathrm{p}\sqrt{\varDelta_\mathrm{p}^2 + 3\delta_\kappa^2} + \frac{3}{2}\delta_\kappa^2 \right]
\end{equation}
and
\begin{equation}
	\delta_\kappa^2 = \delta^2 + \frac{\kappa_1^2}{4}.
\end{equation}

\section{Supplementary Note VII: $S$-parameter background correction and fitting}
\label{sec:Note7}

\subsection{The real transmission function and fit-based background correction}

\begin{figure*}
	\centerline{\includegraphics[width=0.85\textwidth]{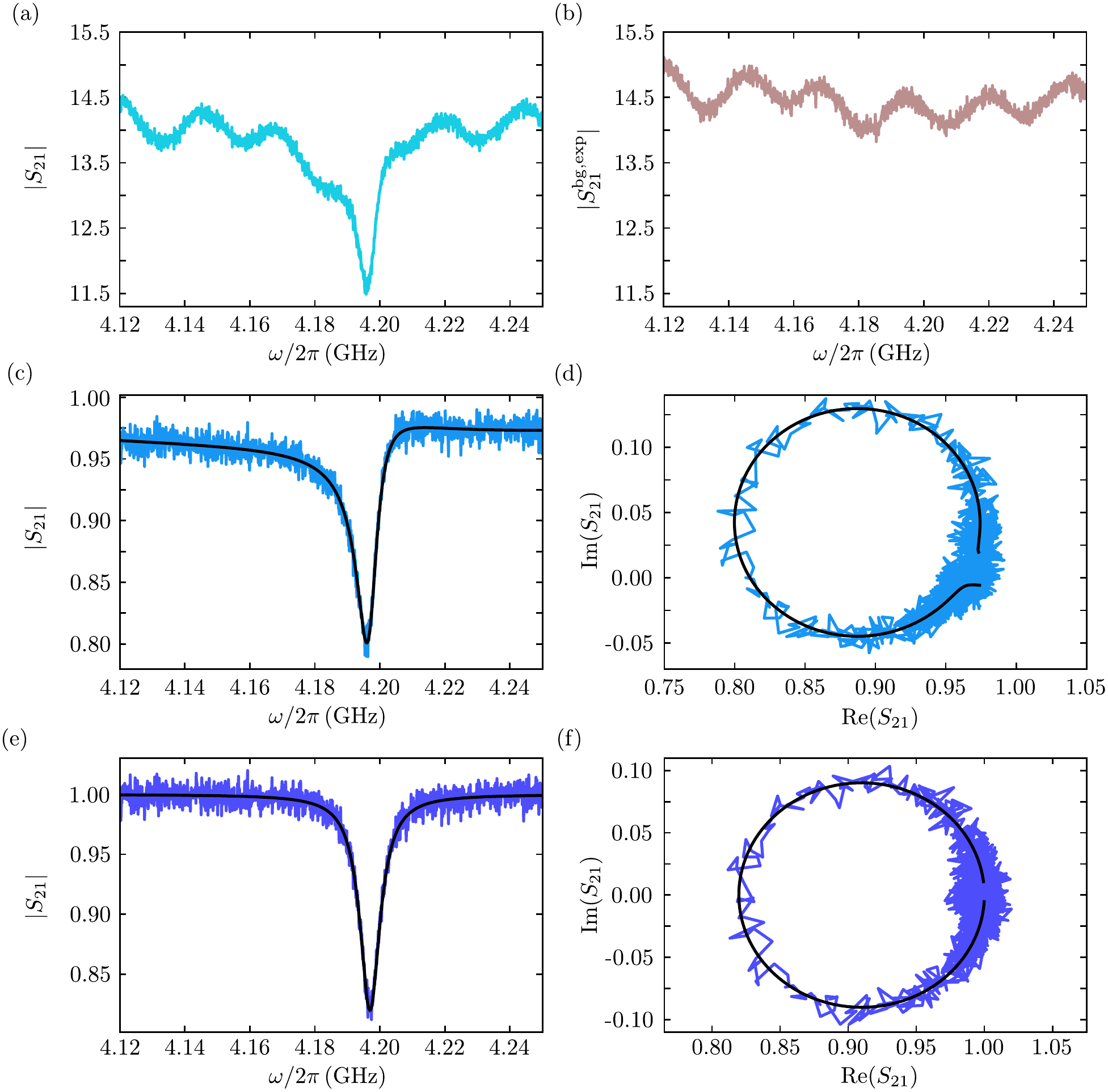}}
	\caption{\textsf{\textbf{Background correction and fitting routine.} (a) Transmission $|S_{21}|$ vs probe frequency of the SQUID circuit 3D$_1$ for a flux bias value close to the flux sweetspot. The absorption resonance dip around $4.197\,$GHz is clearly visible, the measurement temperature is $T_\mathrm{s} = 2.5\,$K. (b) Identical to (a), but at an elevated temperature $T_\mathrm{s} = 3.7\,$K. What we detect here is the experimental background $S_{21}^\mathrm{bg, exp}$, slightly modified by temperature-dependent transmission over the chip and the coldest parts of the microwave cables. We measure not only the amplitude, but also the phase of $S_{21}$ and $S_{21}^\mathrm{bg, exp}$. (c) shows the magnitude of $S_{21}/S_{21}^\mathrm{bg, exp}$, the background is nearly a flat line, but not yet at $|S_{21}| = 1$ as expected for an ideal transmission. (d) shows the imaginary part of the background-divided transmission vs the real part. Noisy light blue lines in (c) and (d) are data, black smooth lines are a fit with Eq.~(\ref{eqn:S21real}). (e) and (f) show the final background-corrected data, where also the remaining background from the fit is divided off and the resonance circle is corrected by the Fano rotation $\theta$. Noisy blue lines in (e) and (f) are data, black smooth lines are the fit.}}
	\label{fig:FigureS6}
\end{figure*}

Due to impedance imperfections in both, the input and output lines, the ideal transmission response is modified by cable resonances and interferences within the setup \cite{Wenner11_SI, Rieger22_SI}.
Origin of these imperfections are connectors, attenuators, wirebonds, transitions to or from the PCB etc. in the signal lines.
In addition, the cabling has a frequency-dependent attenuation.
To take all these modifications into account, we assume that the final transmission parameter $S_{21}^\mathrm{real}$ can be described by
\begin{equation}
	S_{21}^\mathrm{real} = \left(a_0 + a_1\omega + a_2\omega^2  \right)\left[1 - f(\omega)e^{i\theta}\right]e^{i\left(\phi_0  + \phi_1\omega \right)}
	\label{eqn:S21real}
\end{equation}
when the ideal response would be given by
\begin{equation}
	S_{11}^\mathrm{ideal} = 1 - f(\omega).
\end{equation}
The real-valued numbers $a_0, a_1, a_2, \phi_0, \phi_1$ describe a frequency dependent modification of the background transmission, and the phase factor $\theta$ takes into account possible interferences such as parasitic signals bypassing the transmission along the device itself, for instance around the chip.
The exact form of $f(\omega)$ depends on the experiment performed as described above.
Our standard fitting routine begins with removing the actual resonance signal from the transmission, leaving us with a gapped background transmission, which we fit using
\begin{equation}
	S_{21}^\mathrm{bg} = \left(a_0 + a_1\omega + a_2\omega^2  \right)e^{i\left(\phi_0  + \phi_1\omega \right)}.
\end{equation}
Subsequently, we remove this background function from all measurement traces by complex division.
The resonance circle rotation angle $\theta$ is then rotated off additionally.
The result of both corrections is what we present as background-corrected data or transmission/response data in all figures.
For the power dependence measurements, we determine the background from the measurement in the linear regime and perform a background correction based on that single linear response line for all powers.

\subsection{Data-based background correction}
\label{Section:raw_data_processing}

As the circuits in our experiments have a rather large linewidth between several MHz and tens of MHz and as the background transmission cannot be described over such a large frequency span with a simple second order polynomial as suggested by Eq.~(\ref{eqn:S21real}), we perform a two-step background correction to obtain as clean $S$-parameters as possible.
The procedure is exemplarily shown for one resonance of device 3D$_1$ in Supplementary Fig.~\ref{fig:FigureS6}.
In the first step, we record for each measurement (e.g. the one in panel (a)) also the resonance-less transmission function as shown in panel (b).
The resonance-less $S_{21}$ is obtained by increasing the sample temperature to about $T_\mathrm{s} = 3.7\,$K, where the resonance frequency is out of the measurement window and $\kappa_\mathrm{i}$ is so large that the resonance is not impacting the data anymore.
Then we perform a complex division of the full $S_{21}$ signal by the bare background signal $S_{21}^\mathrm{bg, exp}$, the result is a resonance with a nearly flat background as shown in (c), the complex-valued version can be seen in (d).
Subsequently, we perform a fit using Eq.~(\ref{eqn:S21real}) from which we obtain a second background function as well as a Fano rotation angle $\theta$.
We divide off the fit-background, again by complex division, and finally rotate the resonance circle by $\theta$ around its anchor point.
The final result including the corresponding fits can be seen in panels (e) and (f).
For the circuits with constrictions, we perform this data processing with all $S_{21}$ spectra used for the data analysis and all shown resonances have been treated this way.
For the data before constriction cutting, we do only the fit-based background-correction.

\section{Supplementary Note VIII: Additional data and analyses}
\label{sec:Note8}
\subsection{Impact of junction cutting in devices 2D and 3D$_2$}
\label{sec:impact_JJ}
In main paper Fig.~1(f), we present the resonances of device 3D$_1$ before and after cutting the nano-constrictions at $T_\mathrm{s}=2.5\,$K.
From the resonance frequencies before and after cutting, $\omega_\mathrm{b}$ and $\omega_0$, respectively, we determine the inductance of a single constriction $L_\mathrm{c}$ in the device via
\begin{equation}
	L_\mathrm{c} = 2L\left(\frac{\omega_\mathrm{b}^2}{\omega_0^2} - 1\right).
\end{equation}
In Supplementary Fig.~\ref{fig:FigureS7} we show the analogous data for the other two devices of this work, circuits 2D and 3D$_2$.

\begin{figure*}
	\centerline{\includegraphics[width=0.85\textwidth]{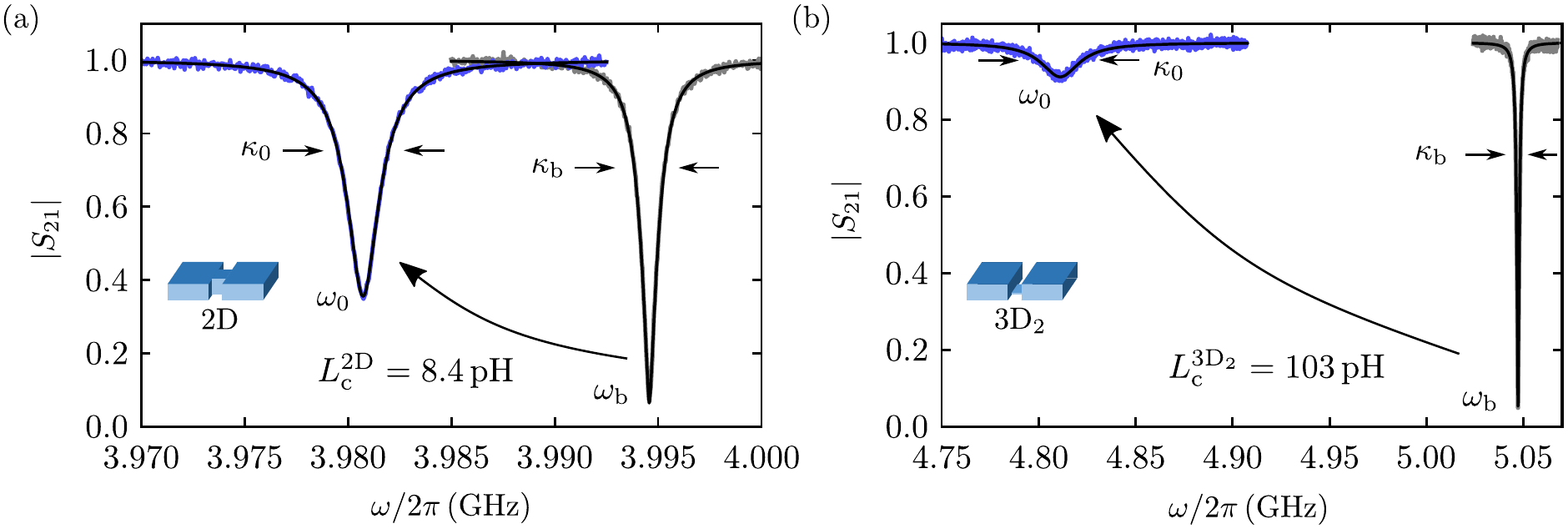}}
	\caption{\textsf{\textbf{Impact of constriction cutting in circuits 2D and 3D$_2$.} (a) Transmission $|S_{21}|$ vs probe frequency of circuit 2D before junction cutting (right, gray noisy line) and after junction cutting at the flux sweetspot (left, blue noisy line). (b) Equivalent of (a), but for circuit 3D$_2$. Both panels: Noisy lines are data, black smooth lines are fits. Measurement temperature $T_\mathrm{s} = 2.5\,$K. Resonance frequency and linewidth before cutting are $\omega_\mathrm{b}$ and $\kappa_\mathrm{b}$, after cutting $\omega_0$ and $\kappa_0$. From the shift, we determine the additional inductance in the circuit $L_\mathrm{c}/2$ (two constrictions in parallel). Values for $\omega_\mathrm{b}, \kappa_\mathrm{b}, \omega_0, \kappa_0$ can be found in the text.}}
	\label{fig:FigureS7}
\end{figure*}
The resonance frequency of 2D has shifted by the cutting from $\omega_\mathrm{b} = 2\pi \cdot 3.995\,$GHz to $\omega_0 = 2\pi \cdot 3.981\,$GHz, which corresponds to a constriction inductance $L_\mathrm{c}^\mathrm{2D} \approx 8.4\,$pH.
The total linewidth has increased from $\kappa_\mathrm{b} = 2\pi\cdot 1.5\,$MHz to $\kappa_0 = 2\pi\cdot 2.4\,$MHz.
In device 3D$_2$ the impact of the shift was much larger, the resonance frequency shifted from $\omega_\mathrm{b} = 2\pi \cdot 5.047\,$GHz to $\omega_0 = 2\pi \cdot 4.811\,$GHz, which corresponds to $L_\mathrm{c}^\mathrm{3D2} \approx 103\,$pH.
The linewidth increased from $\kappa_\mathrm{b} = 2\pi\cdot 2.3\,$MHz to $\kappa_0 = 2\pi\cdot 24.2\,$MHz.
The observation that the linewidths $\kappa$ increase by more after the cutting the larger the constriction inductance is, is not surprising.
As we have shown in Fig.~3 of the main paper, the critical temperature of the constrictions is more suppressed for thinner constrictions and so at the fixed temperature $2.5\,$K the thinner constrictions presumably have a higher thermal quasiparticle density and at the same time a higher microwave current density. 
At this point we cannot exclude that there are also other mechanisms at play such as normal conducting contributions at the surfaces/edges of the constrictions, that increase with neon ion milling time, but the overall trend is understandable quite intuitively from the critical temperature suppression.
Below, we will also discuss the dependence of the linewidths from flux though the SQUIDs.
\subsection{Flux-tuning curves vs sample temperature}

\begin{figure*}
	\centerline{\includegraphics[width=0.85\textwidth]{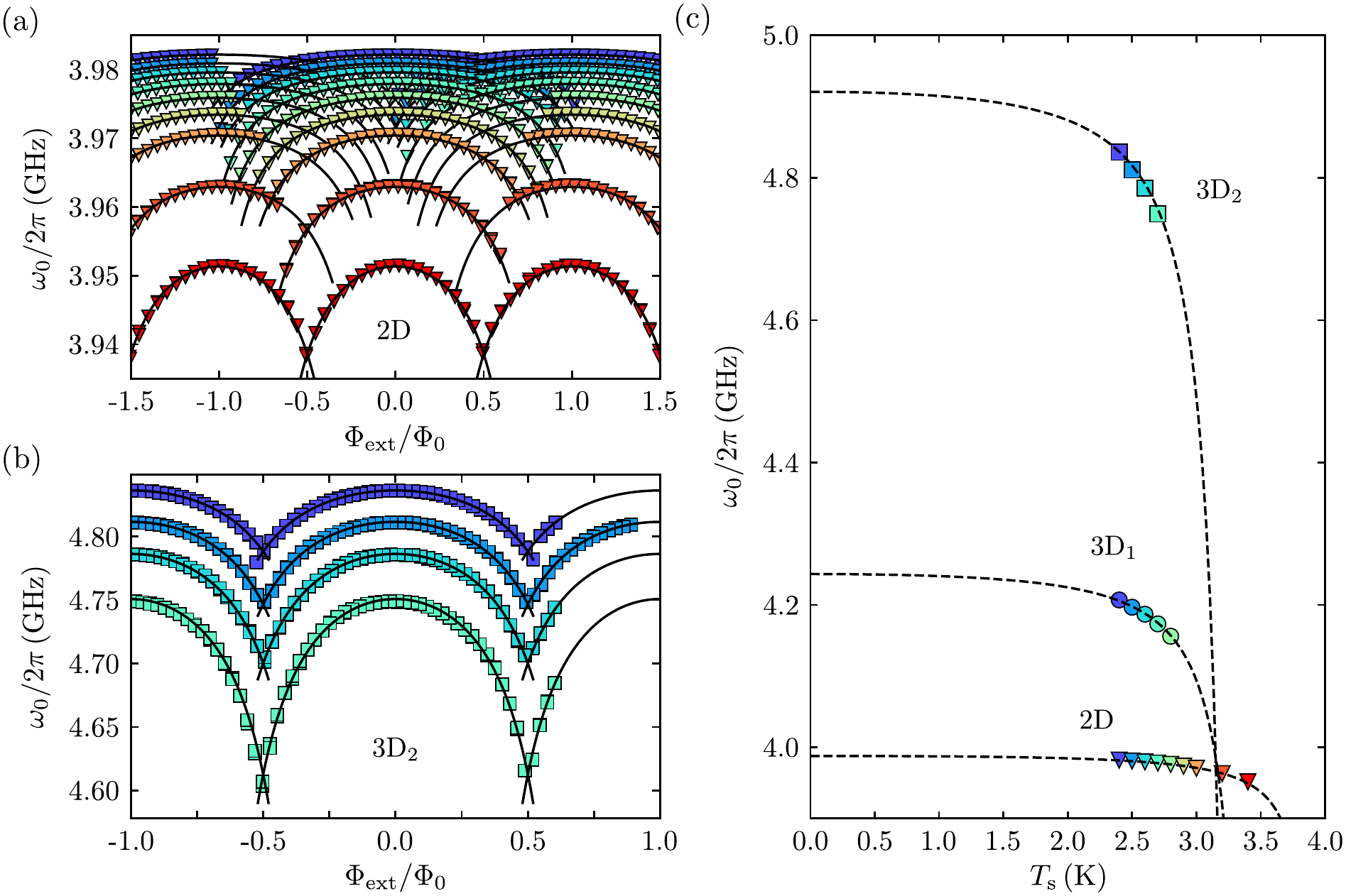}}
	\caption{\textsf{\textbf{Temperature dependence of the flux-tuning curves and the sweetspot resonance frequencies.} (a) shows the resonance frequency vs external bias flux $\omega_0(\Phi_\mathrm{ext})$ in device 2D; symbols are data, lines are fits. Data have been taken at temperatures $T_\mathrm{s} = (2.4, 2.5, 2.6, 2.7, 2.8, 2.9, 3.0, 3.2, 3.4)\,$K, temperature increases from top curve to bottom curve. (b) Equivalent of (a), but for device 3D$_2$. Temperatures are $T_\mathrm{s} = (2.4, 2.5, 2.6, 2.7)\,$K. Data are partly incomplete due to hysteretic jumps (device 2D) and due to insufficient flux data ($\Phi_\mathrm{ext}/\Phi_0 > 0.65$ in device 3D$_2$). In (c) we show the resonance frequency  of all three circuits at the sweetspot frequency ($\Phi_\mathrm{ext}=0$) vs temperature $T_\mathrm{s}$. Symbols are data and the dashed lines are calculated theory lines using Eq.~(\ref{eqn:w0_T}).}}
	\label{fig:FigureS8}
\end{figure*}
In main paper Fig.~3(a) we show flux-tuning curves of the resonance frequency $\omega_0(\Phi_\mathrm{ext})$ for varying sample temperature $T_\mathrm{s}$ in circuit 3D$_1$ and derive from those the critical currents $I_0(T_\mathrm{s})$ and the screening parameters $\beta_\mathrm{L}(T_\mathrm{s})$. 
Since we also show $I_0(T_\mathrm{s})$ and $\beta_\mathrm{L}(T_\mathrm{s})$ for the other two devices 2D and 3D$_2$, we present in Supplementary Fig.~\ref{fig:FigureS8}(a) and (b) the corresponding flux-tuning curves for completeness.
For both devices the sweetspot frequency at $\Phi_\mathrm{ext} = 0$ decreases with increasing $T_\mathrm{s}$ due to the increasing constriction inductance.
At the same time the flux-tuning archs get narrower due to a decreasing screening parameter $\beta_\mathrm{L}$ and an increasing inductance participation ratio $L_\mathrm{c}/L$, and the flux tuning range $\omega_0^\mathrm{max} - \omega_0^\mathrm{min}$ increases for the same reasons.
At the highest temperatures, circuit 2D has a tuning range of $\sim 13\,$MHz and circuit 3D$_2$ of $\sim 150\,$MHz.
From the fits to the data, we extract the screening parameter $\beta_\mathrm{L}$ and the linear constriction contribution $L_\mathrm{lin}$ and calculate the critical current $I_0$.
Screening parameter $\beta_\mathrm{L}$ and critical current $I_0$ are shown in main paper Fig.~3(b) and (c), the linear inductance contributions $L_\mathrm{lin}$ for all three circuits are shown and discussed in the next subsection.
Since the linear inductances in the circuit have a kinetic contribution, the temperature-dependent resonance frequency at the flux sweetspot can be written as
\begin{equation}
	\omega_\mathrm{0}(T_\mathrm{s}) = \frac{1}{\sqrt{C_\mathrm{tot}\left[(L\left(T_\mathrm{s}\right) + L_\mathrm{J0}\left(T_\mathrm{s}\right) + L_\mathrm{lin}\left(T_\mathrm{s}\right)\right]}},
	\label{eqn:w0_T}
\end{equation}
where $L_\mathrm{J0} = \Phi_0 \cdot \left[2\pi I_0(T_\mathrm{s})\right]^{-1}$ and $L_\mathrm{lin}(T_\mathrm{s})$ is discussed in the next subsection.
In Supplementary Fig.~\ref{fig:FigureS8}(c), we show the resonance frequency at the sweetspot of all three devices.
\subsection{The linear inductance contribution $L_\mathrm{lin}$ for all three devices}
The linear inductance contribution $L_\mathrm{lin}$ to the total constriction inductance $L_\mathrm{c}$, which is necessary to model our results, is in agreement with many reports regarding the current-phase-relation (CPR) of niobium constriction junctions.
Essentially all experiments to date found a forward-skewed sinusoidal CPR for this type of junctions, and such a CPR is very similar to the CPR of a series combination of a linear inductance and an ideal Josephson inductance \cite{Likharev79_SI}.
In order to take this linear inductance and its temperature dependence into account for an extrapolation of the SQUID and circuit properties to lower temperatures, we analyze and model the data of $L_\mathrm{lin}$ we have from the limited range of $T_\mathrm{s}$.
In Supplementary Fig.~\ref{fig:FigureS9} all extracted values for $L_\mathrm{lin}$ are shown.
For all three devices $L_\mathrm{lin}$ increases with temperature and the absolute values span between a few pH in device 2D and several $10\,$pH in devices 3D$_1$ and 3D$_2$, the values increase with increasing $L_\mathrm{J0}$.
Since the linear contribution is a kinetic inductance (the geometric inductance of a constriction is negligibly small), we model its temperature dependence
\begin{equation}
	L_\mathrm{lin}(T_\mathrm{s}) = L_\mathrm{off} + \frac{L_\mathrm{lin, 0}}{1 - \left( \frac{T_\mathrm{s}}{T_\mathrm{cc}} \right)^4}
\end{equation}
where $L_\mathrm{lin, 0}$ is the kinetic inductance at zero temperature, $T_\mathrm{cc}$ is the constriction critical temperature and with $L_\mathrm{off}$ we allow for a possible temperature-independent offset.
The result is shown in Supplementary Fig.~\ref{fig:FigureS9} as solid lines.
We note, that due to the small temperature range we could measure the model and theory lines are somewhat speculative and need to be tested in the future by further experiments at lower temperatures.
We believe, however, that it is still useful to extrapolate to lower temperatures by making reasonable assumptions about the temperature-dependence of $I_0$, $L_\mathrm{c}$ and $\beta_\mathrm{L}$, mainly based on theory and earlier results on similar systems.
In panel (b), we show $L_\mathrm{lin}/L_\mathrm{c}$ and see a clear trend for a decrease with increasing temperature, which also is in good agreement with previous reports on the CPR of constriction junctions, since a decreasing $L_\mathrm{lin}/L_\mathrm{c}$ indicates a reduced skewedness as also found experimentally for higher temperatures.
Theory lines in (b) are just directly calculated using the fit lines for $I_0(T_\mathrm{s})$ of main paper Fig.~3(b) and that of Supplementary Fig.~\ref{fig:FigureS9}(a).
It is worth mentioning that it seems that for low temperatures $T_\mathrm{s}/T_\mathrm{cc} \lesssim 0.9$ the relative linear contribution in the 3D devices is larger than in the 2D device.
The reason behind this observation is currently unclear, but it might be related to a damage of the Nb in the 3D constrictions by the neon ion beam, which for instance leads to a smaller superconducting coherence length and therefore deteriorating the constriction quality.

\begin{figure*}
	\centerline{\includegraphics[width=0.85\textwidth]{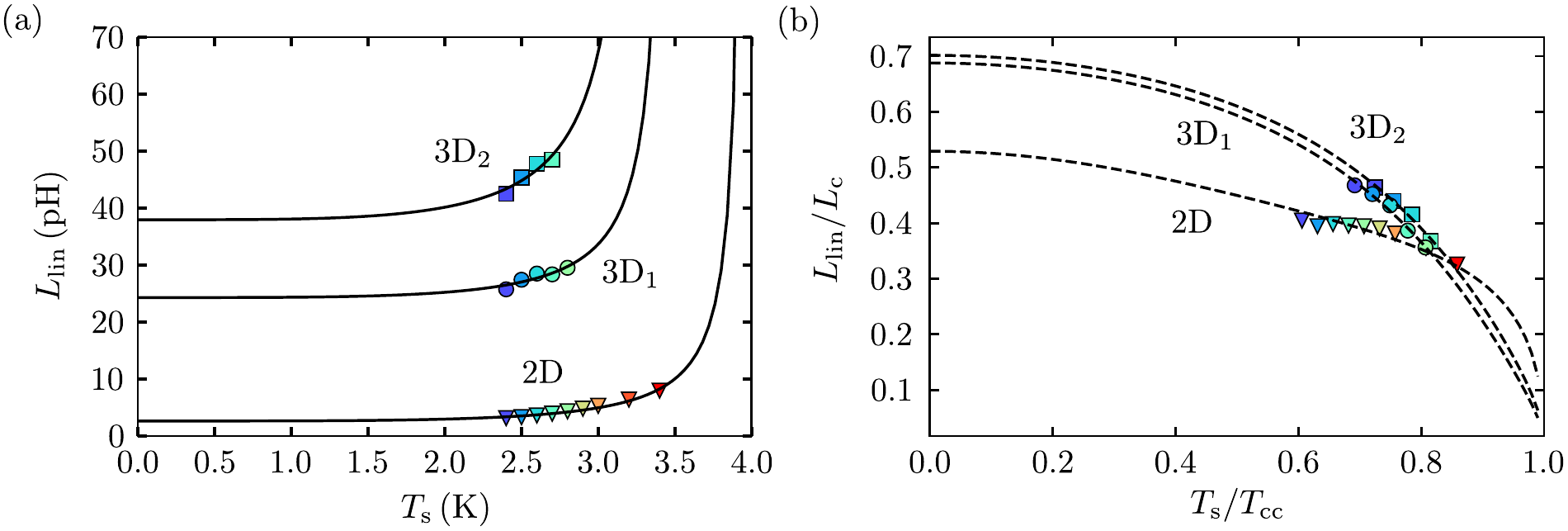}}
	\caption{\textsf{\textbf{Linear contribution $L_\mathrm{lin}$ to the constriction inductance $L_\mathrm{c}$ for all three devices.} (a) Linear constriction inductance contribution $L_\mathrm{lin}$ as obtained from the flux-tuning curve fits vs sample temperature $T_\mathrm{s}$ for all three devices. Symbols are data, lines are fits. The linear inductance increases with decreasing constriction thickness, i.e., with increasing $L_\mathrm{J0}$, and with temperature. (b) Participation ratio of the linear inductance $L_\mathrm{lin}$ to the total constriction inductance $L_\mathrm{c}$ vs reduced temperature $T_\mathrm{s}/T_\mathrm{cc}$, demonstrating that with increasing temperature the linear contribution gets less significant. Symbols are data, dashed lines are calculated from the individual fits of $L_\mathrm{lin}(T_\mathrm{s})$ and $I_0(T_\mathrm{s})$.}}
	\label{fig:FigureS9}
\end{figure*}
\begin{figure*}
	\centerline{\includegraphics[width=0.95\textwidth]{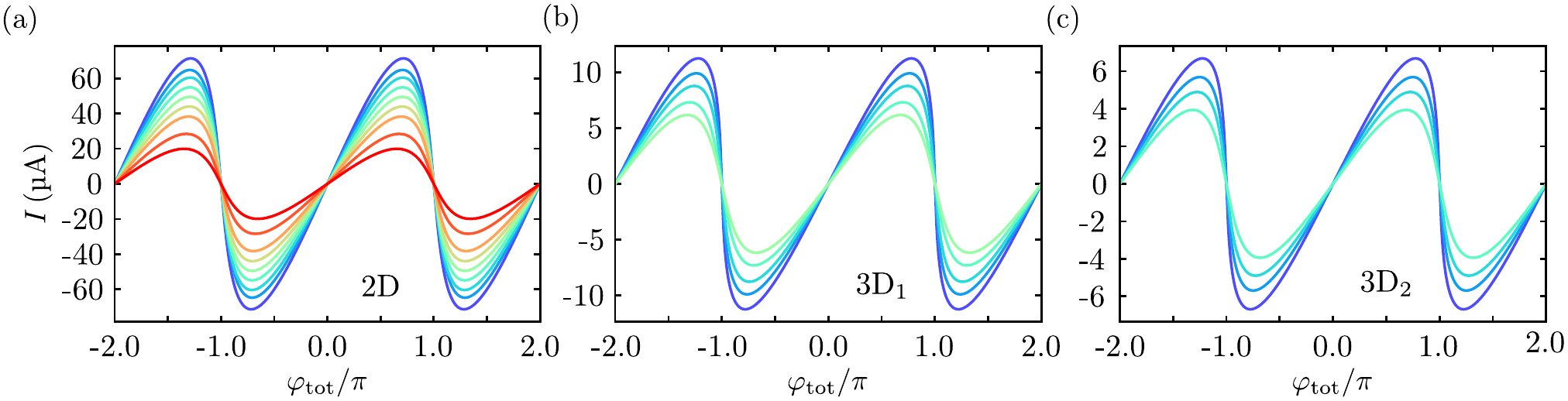}}
	\caption{\textsf{\textbf{Inferred current-phase-relations (CPRs) of all constrictions and all presented temperatures.} CPRs for all three devices at all measurement temperatures $T_\mathrm{s}$, calculated via $L_\mathrm{lin}$ and $I_0$, details cf. text. (a) circuit 2D at $T_\mathrm{s} = (2.4, 2.5, 2.6, 2.7, 2.8, 2.9, 3.0, 3.2, 3.4)\,$K, (b) circuit 3D$_1$ at $T_\mathrm{s} = (2.4, 2.5, 2.6, 2.7, 2.8)\,$K and (c) circuit 3D$_2$ at $T_\mathrm{s} = (2.4, 2.5, 2.6, 2.7)\,$K. Increasing $T_\mathrm{s}$ corresponds to decreasing maximum supercurrent $I_0$ in all panels. All constrictions show the expected forward-skewed CPRs and the skewedness decreases with increasing temperature, which can be seen from the maximum of the curves shifting to smaller phases, when $T_\mathrm{s}$ is increased.}}
	\label{fig:FigureS10}
\end{figure*}

\subsection{The inferred current-phase-relation for all three constriction types}
In order to visualize how the current-phase-relation (CPR) of the three constriction types is looking, which is consistent with our experimental findings and our data analyses, we infer a CPR from the model of an ideal Josephson inductance $L_\mathrm{J0}$ in series with a linear contribution $L_\mathrm{lin}$ for our three circuits and the temperatures we present in the other parts of the manuscript.
To (piecewise) calculate and plot the CPRs, we use for the total phase $\varphi_\mathrm{tot}$ as function of the current $I \leq I_0$
\begin{eqnarray}
	\varphi_\mathrm{tot} & = & \varphi_\mathrm{J} + \varphi_\mathrm{lin} \nonumber \\
	& = & (-1)^n \arcsin{\left( \frac{I}{I_0}\right)} + \frac{2\pi}{\Phi_0}L_\mathrm{lin}I + n\pi
\end{eqnarray}
the result is shown in Supplementary Fig.~\ref{fig:FigureS10}.
All three constriction types show the expected forward-skewed CPRs and the skewedness decreases with increasing temperature, which can be seen from the maximum of the curves shifting to smaller phases, when $T_\mathrm{s}$ is increased.

\subsection{Flux responsivity $\partial\omega_0/\partial\Phi_\mathrm{ext}$ for all devices vs temperature}
Next, we determine the flux responsivity of the three SQUID circuits.
The flux responsivity is the derivative of the resonance frequency with respect to external flux $\partial\omega_0/\partial\Phi_\mathrm{ext}$, usually given in Hz$/\Phi_0$.
This responsivity is an extremely relevant parameter for instance for flux-mediated optomechanics and magnetomechanics \cite{Shevchuk17_SI, Rodrigues19_SI, Schmidt20_SI, Zoepfl20_SI} or for photon-pressure circuits \cite{Johansson14_SI, Eichler18_SI, Bothner21_SI, Rodrigues21_SI}, since it determines by how much the resonance frequency is fluctuating (in first order) for a given amount of flux fluctuations in the SQUID.
In turn, this determines the single-photon coupling rates to the mechanical oscillator for example.
In Supplementary Fig.~\ref{fig:FigureS11} we plot the derivative of the flux-tuning fit curves for all three devices and at all recorded temperatures.
\begin{figure*}
	\centerline{\includegraphics[width=0.95\textwidth]{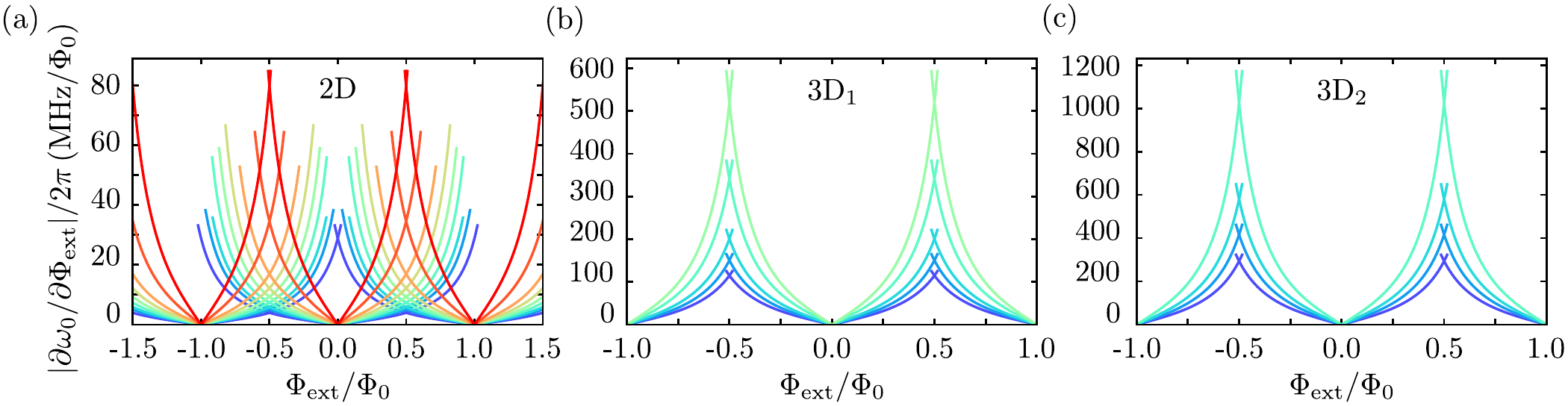}}
	\caption{\textsf{\textbf{Flux responsivity of all three circuits for varying sample temperatures.} Magnitude of the flux responsivity $\partial\omega_0/\partial\Phi_\mathrm{ext}$ vs external flux bias $\Phi_\mathrm{ext}$ for (a) circuit 2D, (b) circuit 3D$_1$ and (c) circuit 3D$_2$. The values are obtained by numerically calculating the derivative of the flux-tuning fit curves for all devices and all temperatures. The sample temperature $T_\mathrm{s}$ is color-coded into the lines, from dark blue $T_\mathrm{s} = 2.4\,$K to red $T_\mathrm{s} = 3.2\,$K. A list of the exact temperatures can be found in the caption of Supplementary Fig.~\ref{fig:FigureS8}.}}
	\label{fig:FigureS11}
\end{figure*}
Not surprisingly and as expected already from Supplementary Fig.~\ref{fig:FigureS8}, we observe that the responsivities increase with increasing temperature and increase with decreasing constriction thickness.
For the 2D device, we find maximum responsivities of about $\gtrsim 80\,$MHz$/\Phi_0$ for the highest temperatures, while for the 3D$_2$ circuit we get up to $\gtrsim 1\,$GHz$/\Phi_0$.
Even for the lower temperatures, we get still several hundred MHz$/\Phi_0$ for the 3D circuits.
The numbers are highly promising for many applications of these circuits, since many other devices based on Aluminum and implemented in recent experiments have still smaller responsivities, but already led to very high radiation-pressure coupling rates to mechanical oscillators or radio-frequency circuits in the MHz domain \cite{Rodrigues21_SI, Bothner22_SI}.

\subsection{Internal linewidth $\kappa_\mathrm{i}$ for all devices vs temperature and flux}

\begin{figure*}
	\centerline{\includegraphics[width=0.85\textwidth]{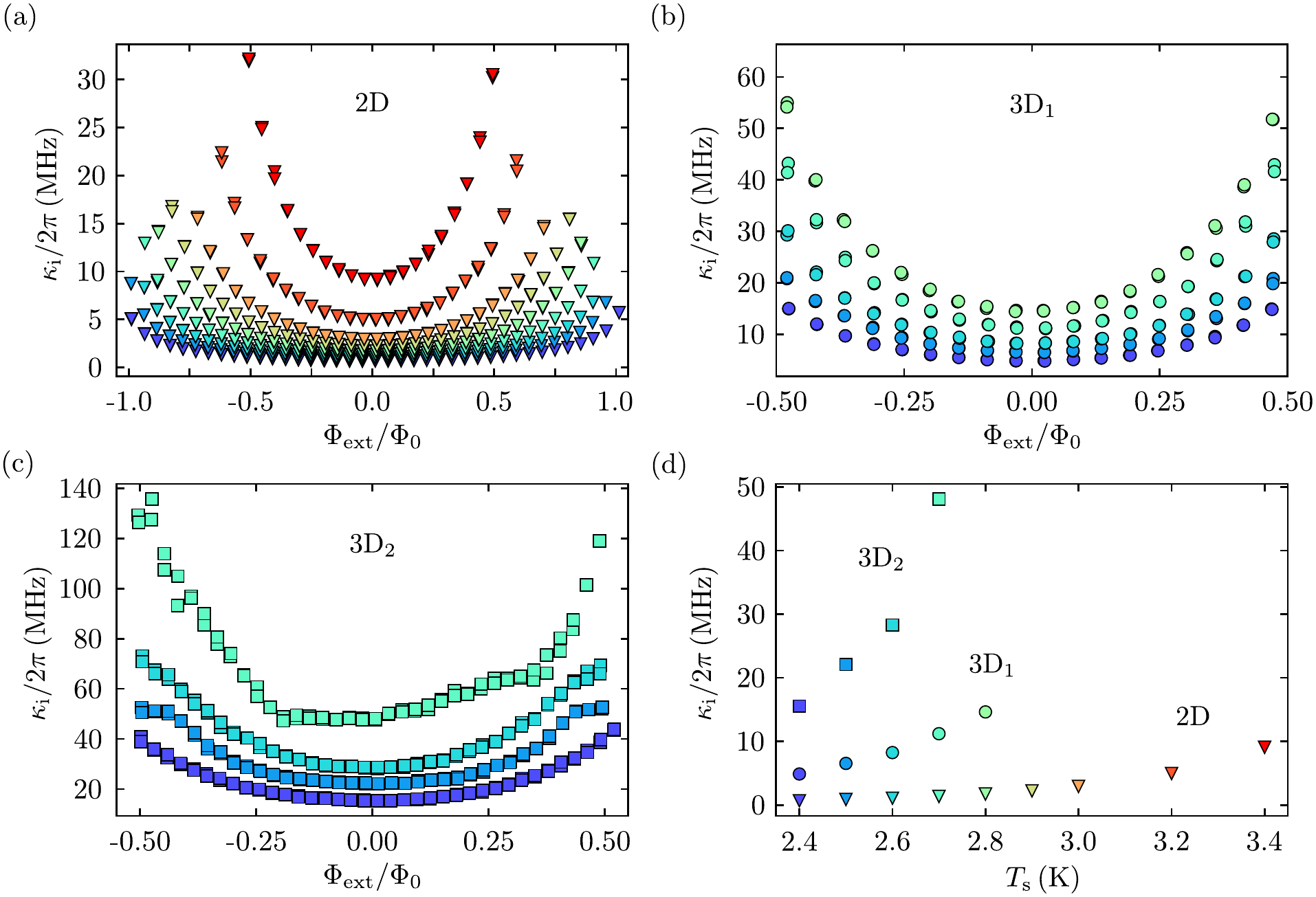}}
	\caption{\textsf{\textbf{Internal linewidth $\kappa_\mathrm{i}$ for all three ciruits for varying sample temperatures.} Internal linewidth $\kappa_\mathrm{i}$ vs external flux bias $\Phi_\mathrm{ext}/\Phi_0$ for (a) circuit 2D, (b) circuit 3D$_1$ and (c) circuit 3D$_2$ at all measurement temperatures $T_\mathrm{s}$. All circuits show an increase of $\kappa_\mathrm{i}$ with increasing flux $|\Phi_\mathrm{ext}/\Phi_0|$ and with increasing sample temperature $T_\mathrm{s}$. Temperatures are equal to the ones in Supplementary Fig.~\ref{fig:FigureS10}. Panel (d) shows $\kappa_\mathrm{i}$ at the sweetspot ($\Phi_\mathrm{ext}/\Phi_0$ = 0) vs sample temperature $T_\mathrm{s}$.}}
	\label{fig:FigureS12}
\end{figure*}
For several experiments as radiation-pressure experiments, parametric amplifiers or dispersive SQUID magnetometry, it is of interest to analyze how the external flux bais affects the losses in the circuit.
Therefore we extract the internal linewidth $\kappa_\mathrm{i}$ for all three circuits for different flux bias points $\Phi_\mathrm{ext}/\Phi_0$ and temperatures $T_\mathrm{s}$, cf Supplementary Fig.~\ref{fig:FigureS12}(a)-(c).
All three devices show a strong dependence of $\kappa_\mathrm{i}$ as function of $\Phi_\mathrm{ext}/\Phi_0$ and show an increase in their linewidth range $\kappa_\mathrm{i}^\mathrm{max} - \kappa_\mathrm{i}^\mathrm{min}$ with increasing temperature.
At the lowest temperature, circuit 2D has a tuning range of $\sim 5\,$MHz, which increases up to $\sim 23\,$MHz for highest sample temperature.
For circuit 3D$_1$ it increases from $\sim 10\,$MHz to $\sim 40\,$MHz and for circuit 3D$_2$ from $\sim 28\,$MHz to $\sim 88\,$MHz.
We believe the increase of internal loss rate with both flux and temperature is related to a locally reduced superconducting energy gap and therefore an increased quasiparticle density in the constriction, mainly due to a reduced critical temperature compared to the rest of the niobium film.
This effect will be enhanced by the external flux bias, since the circulating current through the constrictions is further reducing the gap and increasing the quasiparticle density.
In Supplementary Fig.~\ref{fig:FigureS12}(d) we show the increase of $\kappa_\mathrm{i}$ at the sweetspot ($\Phi_\mathrm{ext}/\Phi_0 = 0$) as function of temperature related to the losses by thermal quasiparticles in the constriction.
It is obvious that the linewidth trends towards smaller values for lower temperatures in all devices.
It will be interesting to analyze the losses at much lower temperatures in future experiments, since currently we do not have a solid model to understand the temperature and flux dependence of $\kappa_\mathrm{i}$ and therefore also cannot make predictions for if and how the losses will saturate in the mK regime.
We show in Supplementary Note~\ref{sec:error_kerr}, that the external linewidth is only weakly dependent on $\Phi_\mathrm{ext}/\Phi_0$ compared to $\kappa_\mathrm{i}$ and this effect is most likely to a slightly different input impedance of the feedline for different resonance frequencies.
The effect is small compared to the change of internal decay rate though.

\section{Supplementary Note IX: Error bars}
\label{sec:Note9}

\subsection{Error bars for the determination of the Kerr constant $\mathcal{K}$}
\label{sec:error_kerr}
\begin{figure*}
	\centerline{\includegraphics[width=0.85\textwidth]{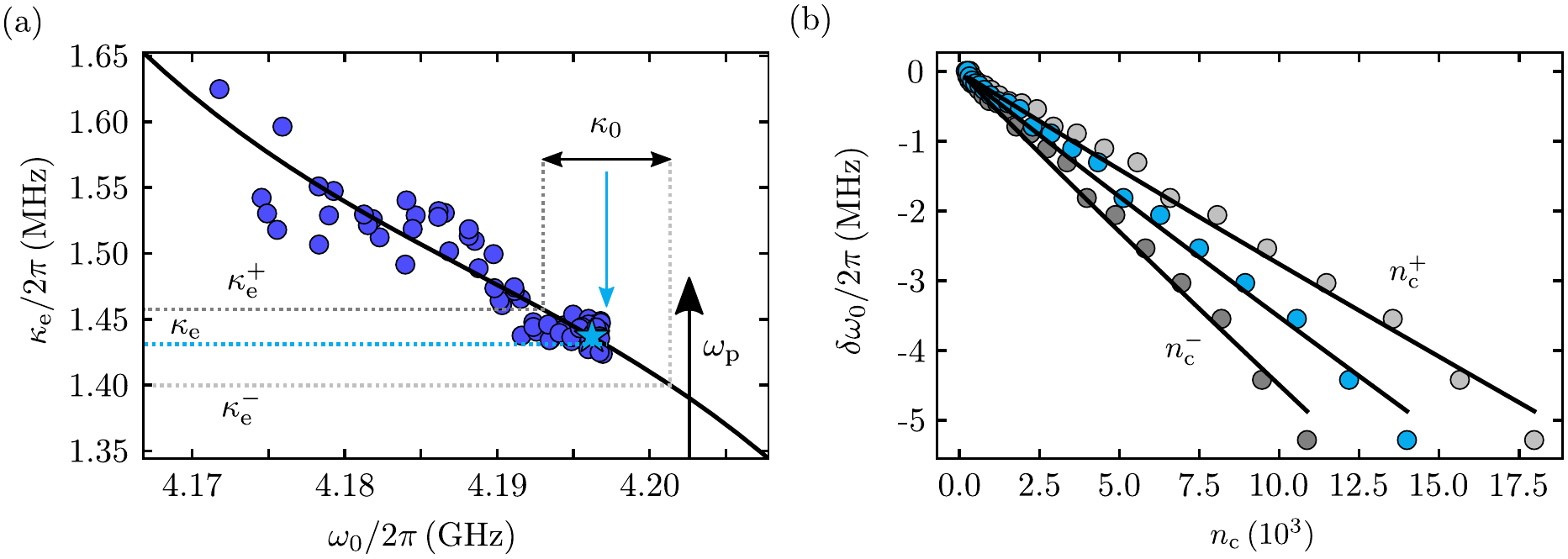}}
	\caption{\textsf{\textbf{Error bar calculation using the uncertainties in $\kappa_\mathrm{e}$ and $P_\mathrm{p}$.} (a) Experimentally determined external linewidth $\kappa_\mathrm{e}/2\pi$ of circuit 3D$_1$ vs circuit resonance frequency $\omega_0/2\pi$ at $T_\mathrm{s}=2.5\,$K. Circles are data, line is a polynomial fit. We describe our procedure for finding $\kappa_\mathrm{e}$ and its uncertainty here exemplarily for a flux bias point $\Phi_\mathrm{ext}/\Phi_0=0.14$. This particular data point is displayed as a star. The pump tone is applied blue-detuned from the resonance frequency at $\omega_\mathrm{p}$ as indicated by the black vertical arrow here. In this case we do not have an experimental value for $\kappa_\mathrm{e}$ at the pump frequency and therefore use the closest available one indicated by the light blue arrow. If we have a value sitting exactly at the pump frequency we use that one. Since $\kappa_\mathrm{e}$ is a function of frequency but our fitting routine is assuming a frequency-independent $\kappa_\mathrm{e}$, we estimate the uncertainty of $\kappa_\mathrm{e}$ to be given by the maximum and minimum of all $\kappa_\mathrm{e}$-values in a window $\kappa_0$ around $\omega_\mathrm{p}$ or the point we chose for $\kappa_\mathrm{e}$, respectively. The window for the particular point chosen here is indicated by the vertical black double-arrow and the corresponding vertical dotted lines. We call the resulting max/min values $\kappa_\mathrm{e}^+$ and $\kappa_\mathrm{e}^-$, respectively. Additionally, we take into account a possible uncertainty in the on-chip pump power by $\pm1\,$dB. As a result, we can find the standard $n_\mathrm{c}$, the maximally possible $n_\mathrm{c}^+$ as well as the minimally possible $n_\mathrm{c}^-$. For all three cases, we plot the frequency shift $\delta\omega_0$ at the star-marked point vs intracavity photon number in panel (b). Circles are experimentally obtained data. We perform three individual fits, shown as lines here. The max/min values obtained for $\mathcal{K}$ from these fits plus their fitting errors are plotted as error bars in main paper Fig.~4(d).}}
	\label{fig:FigureS13}
\end{figure*}
The circuit Kerr nonlinearity $\mathcal{K}$ is an important number for high-dynamic-range applications such as parametric amplifiers, radiation-pressure experiments or dispersive SQUID magnetometry.
To reliably determine the value of $\mathcal{K}$ from the pump-induced frequency shift, a good estimate of the intracircuit photon number $n_\mathrm{c}$ is essential.
As discussed above, we use Eq.~(\ref{eqn:nc_nl}) for calculating $n_\mathrm{c}$, but two of the parameters going into that calculation might come with uncertainties and errors, the on-chip input power $P_\mathrm{p}$ and the external linewidth $\kappa_\mathrm{e}$.
Therefore we also perform an error estimation for $\mathcal{K}$ based on estimated inaccuracies of these two parameters.
For the overall data analysis of the two-tone experiment and to obtain $\mathcal{K}$, we perform first raw data processing of the transmission response $S_\mathrm{21}$ for each pump power as described in Supplementary Note~\ref{Section:raw_data_processing}.
Additionally, we cut out the visible strong pump tone with fixed frequency $\omega_\mathrm{p}$ from all measurements.
As next step, we extract $\omega_0$, $\kappa_\mathrm{e}$ and $\kappa_0$ for all flux bias values by our usual fitting routine, which also means that we have (a rough) knowledge of the frequency-dependence of $\kappa_\mathrm{e}$.
Then, we fit $\kappa_\mathrm{e}$ as a function of $\omega_0$ for the whole flux-tuning-range with a fourth order polynomial, cf. Supplemetary Fig.~\ref{fig:FigureS13}(a).
Since we observe that $\kappa_\mathrm{e}$ is frequency-dependent even over a frequency span of $\kappa_0$, we find the maximum and minimum possible value $\kappa_\mathrm{e}^+$ and $\kappa_\mathrm{e}^-$ for each pump power point as discussed in Supplementary Fig.~\ref{fig:FigureS13}(a) and process these two values as higher and lower errors for $n_\mathrm{c}$.
We estimate the input power on the on-chip feedline by the generator output power and a total input attenuation as described in Supplementary Note~\ref{sec:Note3}.
Additionally a 10$\,$dB directional coupler and a 1$\,$dB microwave cable are added to the pump tone to perform the two-tone experiment.
As uncertainty of the attenuation and as consequence also of the input power we assume $\pm 1\,$dB.
Using Eq.~(\ref{eqn:nc_nl}) and the inaccuracies we get a highest and lowest possible intracavity photon number $n^+_\mathrm{c}$ and $n^-_\mathrm{c}$, respectively, for each pump frequency and pump power point, cf. Supplemetary Fig.~\ref{fig:FigureS13}(b), where the measured frequency shift $\delta\omega_0$ is plotted vs all three photon numbers (lowest, standard, highest).
Finally, we perform a fit for all three cases using Eq.~(\ref{eqn:ttshift}) to determine $\mathcal{K}$, $\mathcal{K}^+$ and $\mathcal{K}^-$.
Note that $|\mathcal{K}^-| > |\mathcal{K}| > |\mathcal{K}^+|$.
The error bars in $\mathcal{K}$ resulting out of the fit with $n^+_\mathrm{c}$, $n^-_\mathrm{c}$ and $n_\mathrm{c}$ are then given by  $|\mathcal{K}^-| - |\mathcal{K}|$  and $|\mathcal{K}| - |\mathcal{K}^+|$ plus their direct errors obtained by the fit.
\section{Supplementary Note X: Calculation of the Kerr nonlinearity}
\label{sec:NoteX}

For the calculation of the Kerr nonlinearity $\mathcal{K}$ we follow the method described in Ref.~\cite{Frattini18_SI} and start with the effective one-dimensional potential for the SQUID, in which each SQUID arm is considered individually
\begin{equation}
	U = \frac{1}{2}E_\mathrm{arm}\left(\varphi_\mathrm{left} -\varphi_1 \right)^2 + \frac{1}{2}E_\mathrm{arm}\left(\varphi_\mathrm{right} -\varphi_2 \right)^2 - E_\mathrm{J}\cos{\varphi_1} - E_\mathrm{J}\cos{\varphi_2}.
\end{equation}
Here $\varphi_1, \varphi_2$ are the phase differences of the two Josephson junctions, $\varphi_\mathrm{left}$ and $\varphi_\mathrm{right}$ are the total phase differences of the left half and the right half of the SQUID loop including the JJs, and the energies are given by
\begin{equation}
	E_\mathrm{J} = \frac{\Phi_0 I_0}{2\pi}, ~~~~~ E_\mathrm{arm} = \frac{\Phi_0^2}{4\pi^2 L_\mathrm{arm}}
\end{equation}
with $L_\mathrm{arm} = \frac{L_\mathrm{loop}}{2} + L_\mathrm{lin}$.
From fluxoid quantization in the SQUID it follows that
\begin{equation}
	\varphi_\mathrm{right} - \varphi_\mathrm{left} = \varphi_\mathrm{ext}
\end{equation}
where $\varphi_\mathrm{ext} = 2\pi\frac{\Phi_\mathrm{ext}}{\Phi_0}$ is the phase introduced by the external flux.
\begin{figure*}[h!]
	\centerline{\includegraphics[width=0.55\textwidth]{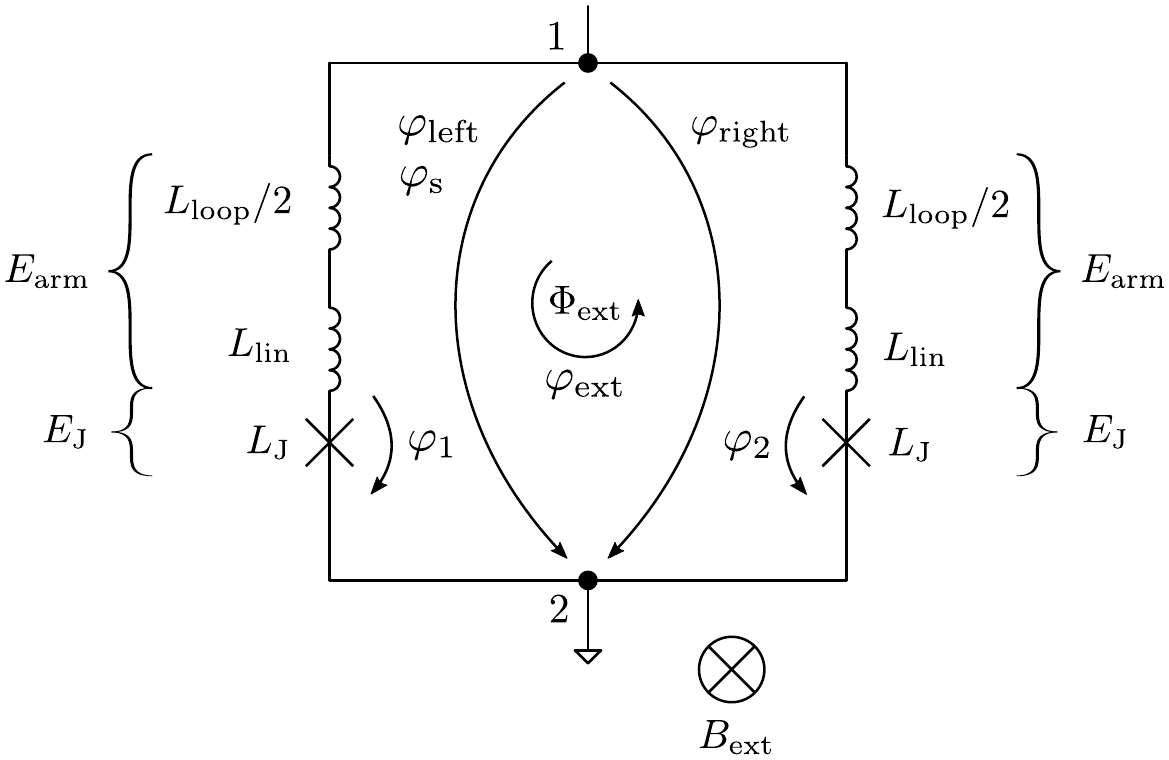}}
	\caption{\textsf{\textbf{Schematic SQUID circuit.} Circuit equivalent of the SQUID with a linear loop inductance $L_\mathrm{loop}/2$ in each arm and a constriction modeled by a linear inductance $L_\mathrm{lin}$ and a sinusoidal Josephson contribution $L_\mathrm{J}$. In each arm the energy $E_\mathrm{J} + E_\mathrm{arm}$ is stored, where $E_\mathrm{J}$ is the Josephson energy and $E_\mathrm{arm}$ is the energy stored in the linear contribution. The nodes 1 and 2 subdivide the loop into a left and right side with a total phase difference $\varphi_\mathrm{left}$ and $\varphi_\mathrm{right}$ between the nodes. The single phase-variable $\varphi_\mathrm{s}$ is identical to $\varphi_\mathrm{left}$. The Josephson contribution of the arms have a phase difference $\varphi_\mathrm{1}$ and $\varphi_\mathrm{2}$, respectively. By applying an external magnetic field $B_\mathrm{ext}$ perpendicular to the SQUID loop an external phase is introduced by the external flux $\Phi_\mathrm{ext}$.}}
	\label{fig:FigureS13}
\end{figure*}
Then, the potential can be written as function of a single phase-variable $\varphi_\mathrm{s} = \varphi_\mathrm{left}$ as
\begin{equation}
	U[\varphi_\mathrm{s}] = \frac{1}{2}E_\mathrm{arm}\left(\varphi_\mathrm{s} -\varphi_1[\varphi_\mathrm{s}] \right)^2 + \frac{1}{2}E_\mathrm{arm}\left(\varphi_\mathrm{s} -\varphi_2[\varphi_\mathrm{s}] - \varphi_\mathrm{ext} \right)^2 - E_\mathrm{J}\cos{\varphi_1[\varphi_\mathrm{s}]} - E_\mathrm{J}\cos{\varphi_2[\varphi_\mathrm{s}]}
\end{equation}
and as boundary conditions we have the current conservation relations \cite{Frattini18_SI}
\begin{eqnarray}
	\varphi_\mathrm{s} & = & \varphi_1 + \zeta\sin{\varphi_1} \label{eqn:phis_phi1}\\
	\varphi_\mathrm{s} & = & \varphi_2 + \zeta\sin{\varphi_2} - \varphi_\mathrm{ext}
	\label{eqn:phis_phi2}
\end{eqnarray}
where $\zeta = E_\mathrm{arm}/E_\mathrm{J} = L_\mathrm{J0}/L_\mathrm{arm}$.
In order to find the Kerr nonlinearity, we have to Taylor-expand the potential up to forth order
\begin{equation}
	\frac{U(\varphi_\mathrm{s})}{E_\mathrm{J}} = c_0 + c_1\left( \varphi_\mathrm{s} -\varphi_\mathrm{s,min}\right) + \frac{c_2}{2}\left( \varphi_\mathrm{s} -\varphi_\mathrm{s,min}\right)^2 + \frac{c_3}{6}\left( \varphi_\mathrm{s} -\varphi_\mathrm{s,min}\right)^3 + \frac{c_4}{24}\left( \varphi_\mathrm{s} -\varphi_\mathrm{s,min}\right)^4 + ...
\end{equation}
where the coefficients are determined by the $n$-th derivative of the potential evaluated at the phase at the potential well minimum $\varphi_\mathrm{s,min}$ 
\begin{equation}
	c_n = \frac{1}{E_\mathrm{J}}\frac{\partial^n U}{\partial\varphi_\mathrm{s}^n}\bigg|_{\varphi_\mathrm{s,min}}.
\end{equation}
To find the value for $\varphi_\mathrm{s,min}$, we demand that in the minimum we have $c_1 = 0$ and as result we get
\begin{equation}
	\varphi_\mathrm{s,min} = \frac{1}{2}\left(\varphi_\mathrm{1, min} + \varphi_\mathrm{2, min} - \varphi_\mathrm{ext}  \right).
	\label{eqn:phimin_condition}
\end{equation}
In the potential minimum, however, i.e., without any phase excitation, we also have
\begin{equation}
	\sin{\varphi_\mathrm{1, min}} = -\sin{\varphi_\mathrm{2, min}}~~~\rightarrow~~~ \varphi_\mathrm{1, min} = -\varphi_\mathrm{2, min}
\end{equation}
since the same SQUID circulating current $J = - I_0\sin{\varphi_1} = I_0\sin{\varphi_2}$ is flowing through both JJs with opposite direction.
Then, using Eq.~(\ref{eqn:phimin_condition}) we can conclude
\begin{equation}
	\varphi_\mathrm{s,min} = -\frac{\varphi_\mathrm{ext}}{2}
\end{equation}
and using Eq.~(\ref{eqn:phis_phi1}) we arrive at
\begin{equation}
	\varphi_\mathrm{1, min} + \zeta\sin{\varphi_\mathrm{1, min} + \frac{\varphi_\mathrm{ext}}{2}} = 0
\end{equation}
which is completely equivalent to
\begin{equation}
	\frac{\Phi}{\Phi_0}	= \frac{\Phi_\mathrm{ext}}{\Phi_0} - \frac{\beta_\mathrm{L}}{2}\sin{\left( \pi\frac{\Phi}{\Phi_0}\right)}
\end{equation}
with the screening parameter $\beta_\mathrm{L}$, the total flux in the SQUID $\Phi$ and when using the relation $\varphi_\mathrm{1,min} = -\pi\frac{\Phi}{\Phi_0}$.
For the derivatives, we get
\begin{eqnarray}
	\frac{\partial U}{\partial\varphi_\mathrm{s}} & = & E_\mathrm{arm}\left(2\varphi_\mathrm{s} -\varphi_1[\varphi_\mathrm{s}] -\varphi_2[\varphi_\mathrm{s}] + \varphi_\mathrm{ext} \right) \\
	\frac{\partial^2 U}{\partial\varphi_\mathrm{s}^2} & = & E_\mathrm{arm}\left( 2 - \frac{\partial\varphi_1}{\partial\varphi_\mathrm{s}} - \frac{\partial\varphi_2}{\partial\varphi_\mathrm{s}} \right) \\
	\frac{\partial^3 U}{\partial\varphi_\mathrm{s}^3} & = & -E_\mathrm{arm}\left( \frac{\partial^2\varphi_1}{\partial\varphi_\mathrm{s}^2} + \frac{\partial^2\varphi_2}{\partial\varphi_\mathrm{s}^2} \right) \\
	\frac{\partial^4 U}{\partial\varphi_\mathrm{s}^4} & = & -E_\mathrm{arm}\left( \frac{\partial^3\varphi_1}{\partial\varphi_\mathrm{s}^3} + \frac{\partial^3\varphi_2}{\partial\varphi_\mathrm{s}^3} \right)
\end{eqnarray}
and the phase derivatives we can obtain from Eqs.~(\ref{eqn:phis_phi1}, \ref{eqn:phis_phi2}) as
\begin{eqnarray}
	\frac{\partial \varphi_\mathrm{s}}{\partial \varphi_1} & = & 1 + \zeta\cos{\varphi_1} \\
	\frac{\partial \varphi_\mathrm{s}}{\partial \varphi_2} & = & 1 + \zeta\cos{\varphi_2} 
\end{eqnarray}
which can be inverted as
\begin{eqnarray}
	\frac{\partial \varphi_1}{\partial \varphi_\mathrm{s}} & = & \frac{1}{1 + \zeta\cos{\varphi_1}} \\
	\frac{\partial \varphi_2}{\partial \varphi_\mathrm{s}} & = & \frac{1}{1 + \zeta\cos{\varphi_2}}.
\end{eqnarray}
The consecutive derivatives are for $j = 1, 2$
\begin{eqnarray}
	\frac{\partial^2 \varphi_j}{\partial\varphi_\mathrm{s}^2} & = & \frac{\zeta\sin{\varphi_j}}{\left(1 + \zeta\cos{\varphi_j} \right)^3} \\
	\frac{\partial^3 \varphi_j}{\partial\varphi_\mathrm{s}^3} & = & \frac{\zeta\cos{\varphi_j}\left( 1 + \zeta\cos{\varphi_j} \right) + 3\zeta^2\sin^2{\varphi_j}}{\left(1 + \zeta\cos{\varphi_j} \right)^5}
\end{eqnarray}
which we can finally use to express our Taylor coefficients with $\varphi_0 = -\varphi_\mathrm{1,min} = \pi\frac{\Phi}{\Phi_0}$ as
\begin{eqnarray}
	c_2 & = & \frac{2\cos{\varphi_0}}{1 + \zeta\cos{\varphi_0}} \\
	c_3 & = & 0 \\
	c_4 & = & -2\frac{\cos{\varphi_0}\left( 1 + \zeta\cos{\varphi_0} \right) + 3\zeta \sin^2{\varphi_0}}{\left(1 + \zeta\cos{\varphi_0} \right)^5}.
\end{eqnarray}
The SQUID inductance and Kerr nonlinearity of the SQUID, when shunted with $C_\mathrm{tot}$ are now given by \cite{Frattini18_SI}
\begin{eqnarray}
	L_\mathrm{s} & = & \frac{L_\mathrm{J0}}{c_2} \nonumber \\
	& = & \frac{1}{2}\left( L_\mathrm{J} + L_\mathrm{arm} \right)
\end{eqnarray}
and
\begin{eqnarray}
	\mathcal{K}_\mathrm{s} & = & \frac{e^2}{2\hbar C_\mathrm{tot}}\frac{c_4}{c_2} \nonumber \\
	& = & -\frac{e^2}{2\hbar C_\mathrm{tot}}\left( \frac{L_\mathrm{J}}{L_\mathrm{arm} + L_\mathrm{J}} \right)^3\left[ 1 + 3\frac{L_\mathrm{arm}}{L_\mathrm{arm} + L_\mathrm{J}}\tan^2{\varphi_0} \right]
\end{eqnarray}
where $L_\mathrm{J} = L_\mathrm{J0}/\cos{\varphi_0}$, $e$ is the elementary charge and $\hbar$ the reduced Planck number.
When we add a linear inductance $L - L_\mathrm{loop}/4$ in series, we get the modified parameters \cite{Frattini18_SI}
\begin{eqnarray}
	\tilde{c}_2 & = & pc_2 \\
	L_\mathrm{tot} & = & \frac{L_\mathrm{s}}{p} \\
	\tilde{c}_4 & = & p^4 c_4 \\
	\mathcal{K} & = & p^3 \mathcal{K}
\end{eqnarray}
where $p$ is the inductance participation ratio
\begin{equation}
	p = \frac{L_\mathrm{s}}{L - L_\mathrm{loop}/4 + L_\mathrm{s}}.
\end{equation}
Then, we have finally the explicit expression for the circuit Kerr nonlinearity
\begin{equation}
	\mathcal{K} = -\frac{e^2}{2\hbar C_\mathrm{tot}}\left( \frac{L_\mathrm{J}}{2L + L_\mathrm{lin} + L_\mathrm{J}} \right)^3\left[ 1 + 3\frac{L_\mathrm{arm}}{L_\mathrm{arm} + L_\mathrm{J}}\tan^2{\varphi_0} \right]
\end{equation}
which we use in main paper Fig.~4.

\let\oldaddcontentsline\addcontentsline
\renewcommand{\addcontentsline}[3]{}

\end{document}